\documentclass[a4,aps,prd,superscriptaddress,floatfix,nofootinbib,showpacs]
{revtex4}

\usepackage{color}
\usepackage[dvips]{graphicx}

\newcommand{\lsim}{\mathrel{\mathop{\kern 0pt \rlap
  {\raise.2ex\hbox{$<$}}}
  \lower.9ex\hbox{\kern-.190em $\sim$}}}
\newcommand{\gsim}{\mathrel{\mathop{\kern 0pt \rlap
  {\raise.2ex\hbox{$>$}}}
  \lower.9ex\hbox{\kern-.190em $\sim$}}}

\newcommand{\be}{\begin{equation}}
\newcommand{\ee}{\end{equation}}
\newcommand{\beqarr}{\begin{eqnarray}}
\newcommand{\eeqarr}{\end{eqnarray}}

\begin{document}

\preprint{DFTT 3/2008, KIAS--P08014}

\title{
\vspace*{-1cm}\hfill\hspace{4.1in}{\small DFTT 3/2008, KIAS--P08014} \\
\vspace*{1cm}
Light neutralinos at LHC in cosmologically--inspired scenarios:
new benchmarks in the search for supersymmetry}



\author{A. Bottino}
\affiliation{Dipartimento di Fisica Teorica, Universit\`a di Torino \\
Istituto Nazionale di Fisica Nucleare, Sezione di Torino \\
via P. Giuria 1, I--10125 Torino, Italy}

\author{N. Fornengo}
\affiliation{Dipartimento di Fisica Teorica, Universit\`a di Torino \\
Istituto Nazionale di Fisica Nucleare, Sezione di Torino \\
via P. Giuria 1, I--10125 Torino, Italy}

\author{G. Polesello}
\affiliation{Istituto Nazionale di Fisica Nucleare, Sezione di Pavia \\
 via Bassi 6, 27100 Pavia Italy}

\author{S. Scopel}
\affiliation{Korea Institute for Advanced Study \\
 Seoul 130-722, Korea}

\date{\today}

\begin{abstract}
\vspace{1cm} We study how the properties of the four neutralino states,
$\chi_i$ (i = 1, 2, 3, 4), can be investigated at the Large Hadron
Collider (LHC), in the case when the lightest one, $\chi_1$, has a
mass $m_{\chi} \lsim 50$ GeV and is stable. This situation arises
naturally in supersymmetric models where gaugino masses are not
unified at a Grand Unified (GUT) scale and R-parity is conserved.  The
main features of these neutralino states are established by
analytical and numerical analyses, and two scenarios are singled out on
the basis of the cosmological properties required for the relic
neutralinos.  Signals expected at LHC are discussed through the main
chain processes started by a squark, produced in the initial
proton-proton scattering.  We motivate the selection of some
convenient benchmarks, in the light of the spectroscopical properties
(mass spectrum and transitions) of the four neutralino
states. Branching ratios and the expected total number of events are
derived in the various benchmarks, and their relevance for experimental
determination of neutralino properties is finally discussed.
\end{abstract}


\maketitle

\section{Introduction}
\label{sec:intro}

Light neutralinos, {\it i.e.} neutralinos with a mass $m_{\chi}  \lsim 50$ GeV,
arise naturally in supersymmetric models where gaugino masses are not
 unified at a Grand Unified (GUT) scale. When R-parity conservation is assumed,
these neutralinos offer a very rich phenomenology under various
cosmological and astrophysical aspects \cite{lowneu,lowdir}.

In the present paper we study how these light neutralinos  can
be investigated at the Large Hadron Collider (LHC), which will soon start
being operated at CERN \cite{atltdr,cmstdr2}.

To this purpose we first delineate what are the
main properties of the four neutralino states, $\chi_i$ (i = 1, 2, 3, 4),
in the case when the lightest one, $\chi_1$ (or $\chi$ in short), has
a mass $m_{\chi}  \lsim 50$ GeV. This neutralino is stable, in case it
occurs to be  the Lightest Supersymmetric Particle and R-parity
is conserved. In particular, special asymptotic schemes for the four
neutralino states are analysed: two hierarchical schemes (with so-called normal and
inverted hierarchy, respectively) and an almost degenerate one.

Once the properties of  all the neutralino states are established
by analytical and numerical analyses, two main scenarios are  singled out  on the
basis of the cosmological properties required for the relic neutralinos.

Then, we analyse the signals expected at LHC, in terms of the main
chain processes which are initiated by a squark and lead finally to
$\chi_1$ through an intermediate production of $\chi_i$ (i = 2, 3,
4). We evaluate the branching ratios for the various processes and
discuss their features in terms of the spectroscopic properties
(mass spectrum and transitions) of the four neutralino states.
Various benchmarks of special physical interest are considered.

The total number of events expected for these processes at LHC is
estimated for a typical representative value of the integrated
luminosity and their relevance for experimental determination of
neutralino properties is finally discussed.

The scheme of our paper is the following. The features of the employed
supersymmetric model are presented in Sect. \ref{sec:susy}, where also
the main structural properties of the four neutralino states are
described. In Sect. \ref{sec:scenarios} we delineate the scenarios
imposed by cosmology on light relic neutralinos. In
Sect. \ref{sec:signals} we derive branching ratios and the total number
of events expected at LHC. Finally, conclusions are drawn in
Sect. \ref{sec:conclusions}.

\section{The supersymmetric model}
\label{sec:susy}

The supersymmetric scheme we employ in the present paper is the
one described in Ref. \cite{lowneu}: an effective MSSM scheme
(effMSSM) at the electroweak scale, with the following independent
parameters: $M_1, M_2, M_3, \mu, \tan\beta, m_A, m_{\tilde q}, m_{\tilde l}$
and $A$. Notations are as
follows: $M_1$,  $M_2$  and $M_3$ are the U(1), SU(2) and SU(3)  gaugino masses 
(these parameters are taken here to be positive),
$\mu$ is the Higgs mixing mass parameter, $\tan\beta$ the
ratio of the two Higgs v.e.v.'s, $m_A$ the mass of the CP-odd
neutral Higgs boson, $m_{\tilde q}$ is a squark soft--mass common
to all squarks, $m_{\tilde l}$ is a slepton soft--mass common to
all sleptons, and $A$ is a common dimensionless trilinear parameter
for the third family, $A_{\tilde b} = A_{\tilde t} \equiv A
m_{\tilde q}$ and $A_{\tilde \tau} \equiv A m_{\tilde l}$ (the
trilinear parameters for the other families being set equal to
zero).
%
%
In our model, no gaugino mass unification at a Grand Unified (GUT)
scale is assumed. The following experimental constraints are imposed:
accelerators data on supersymmetric and Higgs boson searches (CERN
$e^+ e^-$ collider LEP2 \cite{LEPb} and Collider Detectors D0 and CDF
at Fermilab \cite{cdf}); measurements of the $b \rightarrow s +
\gamma$ decay process \cite{bsgamma}: 2.89 $\leq B(b \rightarrow s +
\gamma) \cdot 10^{4} \leq$ 4.21 is employed here (this interval is
larger by 25\% with respect to the experimental determination
\cite{bsgamma} in order to take into account theoretical uncertainties
in the supersymmetric (SUSY) contributions \cite{bsgamma_theorySUSY} to the branching
ratio of the process (for the Standard Model calculation, we employ
the recent NNLO results from Ref.  \cite{bsgamma_theorySM})); the upper
bound on the branching ratio $BR(B_s^{0} \rightarrow \mu^{-} +
\mu^{+})$ \cite{bsmumu}: we take $BR(B_s^{0} \rightarrow \mu^{-} +
\mu^{+}) < 1.2 \cdot 10^{-7}$; measurements of the muon anomalous
magnetic moment $a_\mu \equiv (g_{\mu} - 2)/2$: for the deviation
$\Delta a_{\mu}$ of the experimental world average from the
theoretical evaluation within the Standard Model we use here the range
$-98 \leq \Delta a_{\mu} \cdot 10^{11} \leq 565 $, derived from the
latest experimental \cite{bennet} and theoretical \cite{bijnens} data.

\subsection{Composition of the neutralino states}
\label{sec:susy1}

 The linear superpositions
of bino $\tilde B$, wino $\tilde W^{(3)}$ and of the two Higgsino
states $\tilde H_1^{\circ}$, $\tilde H_2^{\circ}$ which define the
four neutralino states are written in the following way:
\begin{equation}
\chi_i \equiv a_1^{(i)} \tilde B + a_2^{(i)} \tilde W^{(3)} +
a_3^{(i)} \tilde H_1^{\circ} + a_4^{(i)}  \tilde H_2^{\circ} \; \; \; \;
(i=1,2,3,4).
\label{neutralino}
\end{equation}

\noindent
These states diagonalize the mass matrix

\begin{equation}
\left( \begin{array}{cccc} M_1 & 0 & - m_Z s_{\theta} c_{\beta}  & m_Z s_{\theta} s_{\beta} \\
                           0 & M_2 & m_Z c_{\theta} c_{\beta} & - m_Z c_{\theta} s_{\beta} \\
- m_Z s_{\theta} c_{\beta} & m_Z c_{\theta} c_{\beta} & 0 & - \mu \\
m_Z s_{\theta} s_{\beta} & - m_Z c_{\theta} s_{\beta} & - \mu & 0 \label{massmatrix} \end{array}
\right),
\end{equation}

\noindent
where $s_{\beta} \equiv \sin {\beta}$, $c_{\beta} \equiv \cos {\beta}$, and
$s_{\theta} \equiv \sin {\theta}$, $c_{\theta} \equiv \cos {\theta}$,
$\theta$ being the Weinberg angle.
The mass eigenvalues (with signs) will be denoted by $m_i$. The smallest
mass eigenvalue $|m_1|$ will also be denoted by $m_{\chi}$ .

From the set of equations

\begin{eqnarray}
(M_1 - m_i) a_1^{(i)} - m_Z s_{\theta} c_{\beta} a_3^{(i)} + m_Z s_{\theta} s_{\beta} a_4^{(i)} = 0 \nonumber \\
(M_2 - m_i) a_2^{(i)} + m_Z c_{\theta} c_{\beta} a_3^{(i)} - m_Z c_{\theta} s_{\beta} a_4^{(i)} = 0 \nonumber \\
 - m_Z s_{\theta} c_{\beta} a_1^{(i)} + m_Z c_{\theta} c_{\beta} a_2^{(i)} - m_i a_3^{(i)} - \mu  a_4^{(i)}= 0 \nonumber \\
 m_Z s_{\theta} s_{\beta} a_1^{(i)} - m_Z c_{\theta} s_{\beta} a_2^{(i)} - \mu a_3^{(i)} - m_i  a_4^{(i)}= 0,
\label{diagg}
\end{eqnarray}

\noindent
that follow from the diagonalization of the mass matrix of Eq. (\ref{massmatrix}), one obtains
the ratios
\begin{eqnarray}
\frac{a_2^{(i)}}{a_1^{(i)}} &=&  - \frac{M_1 - m_i}{M_2 - m_i} cot \theta \nonumber \\
\frac{a_3^{(i)}}{a_1^{(i)}} &=&   \frac{m_Z(m_i c_\beta + \mu s_\beta)(s_\theta^2 M_2 +
c_\theta^2 M_1 - m_i)}{s_\theta (M_2 - m_i)(\mu^2 - m_i^2)}  \nonumber \\
\frac{a_3^{(i)}}{a_4^{(i)}} &=&  - \frac{m_i c_\beta + \mu s_\beta}
{\mu c_\beta + m_i s_\beta},
\label{diag}
\end{eqnarray}

\noindent
which, together with the normalization conditions,  provide the compositions of the four
neutralino states  \cite{esa}.

The LEP lower limit on the chargino mass ($m_{{\chi}^{\pm}} \gsim$ 100 GeV) sets a lower bound on
both $|\mu|$ and $M_2$: $|\mu|, M_2 \gsim$ 100 GeV. Since, on the contrary, $M_1$ is unbound,
the lowest value for $m_{\chi}$ occurs when

\begin{equation}
m_{\chi} \simeq M_1 << |\mu|, M_2.
\label{approx}
\end{equation}

\noindent
Thus, $\chi \equiv \chi_1$ is mainly a Bino, whose mixings with the other interaction eigenstates are readily
derived from Eqs. (\ref{diag}) to be

 \begin{equation}
\frac{a_2^{(1)}}{a_1^{(1)}} \simeq \frac{\xi_1}{M_2} cot_\theta, \; \;
\frac{a_3^{(1)}}{a_1^{(1)}} \simeq s_\theta s_\beta \frac{m_Z}{\mu}, \; \;
\frac{a_3^{(1)}}{a_4^{(1)}} \simeq  - \frac{\mu s_\beta}{M_1 s_\beta + \mu c_\beta},
\label{approx1}
\end{equation}

\noindent
where $\xi_1 \equiv m_1 - M_1$. Notice that, in deriving
Eq. (\ref{approx}), we have also taken into account that in the
present paper we will only consider $\tan \beta \geq$ 10.

From Eqs. (\ref{diag}) one also obtains approximate expressions for
the compositions of the eigenstates which
correspond to the asymptotic mass eigenvalues: $m_i \sim \pm \mu$ and $m_i \sim M_2$. That is:

a) for the  neutralino states $\chi_i$ with $m_i \simeq \pm \mu$,

\begin{equation}
\frac{a_2^{(i)}}{a_1^{(i)}} \simeq  \frac{\pm \mu}{M_2  \mp \mu} cot_\theta, \; \;
\frac{a_1^{(i)}}{a_3^{(i)}}
\simeq  \frac{2 \xi_2 s_\theta (\pm \mu - M_2)}{M_Z  s_\beta ({s_\theta}^2 M_2  \mp \mu)}, \; \;
\frac{a_3^{(i)}}{a_4^{(i)}} \simeq \mp 1 + \frac{\xi_2}{\mu},
\label{approx23}
\end{equation}

\noindent
where $\xi_2 \equiv \pm \mu -  m_i$.

b) for the  neutralino state $\chi_i$ with $m_i \simeq M_2$,

\begin{equation}
\frac{a_1^{(i)}}{a_2 ^{(i)}} \simeq \frac{\xi_3}{M_2} \tan_\theta, \; \;
\frac{a_1^{(i)}}{a_3^{(i)}} \simeq \frac{\xi_3 s_\theta (M_2^2 - \mu^2)}{M_Z (M_2 c_{\beta} +
\mu s_{\beta}) c_{\theta}^2 M_2}, \; \;
\frac{a_3^{(i)}}{a_4^{(i)}} \simeq  - \frac{\mu s_\beta + M_2 c_\beta}{M_2 s_\beta + \mu c_{\beta}},
\label{approx4}
\end{equation}

\noindent
where $\xi_3 \equiv  M_2 - m_i$.

\begin{figure}
\begin{center}
\includegraphics[width=8cm]{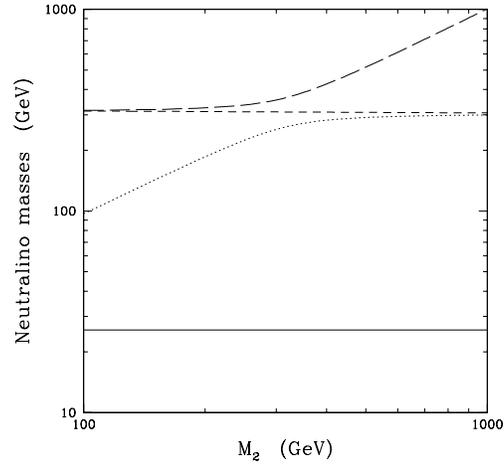}
\end{center}
\caption{Neutralino masses
as functions of $M_2$ for $M_1=25$ GeV,
$\mu=300$ GeV and $\tan\beta=$10.
\label{fig:neutralino_masses}.}
\end{figure}

\begin{figure}
\begin{center}
\includegraphics[width=12cm]{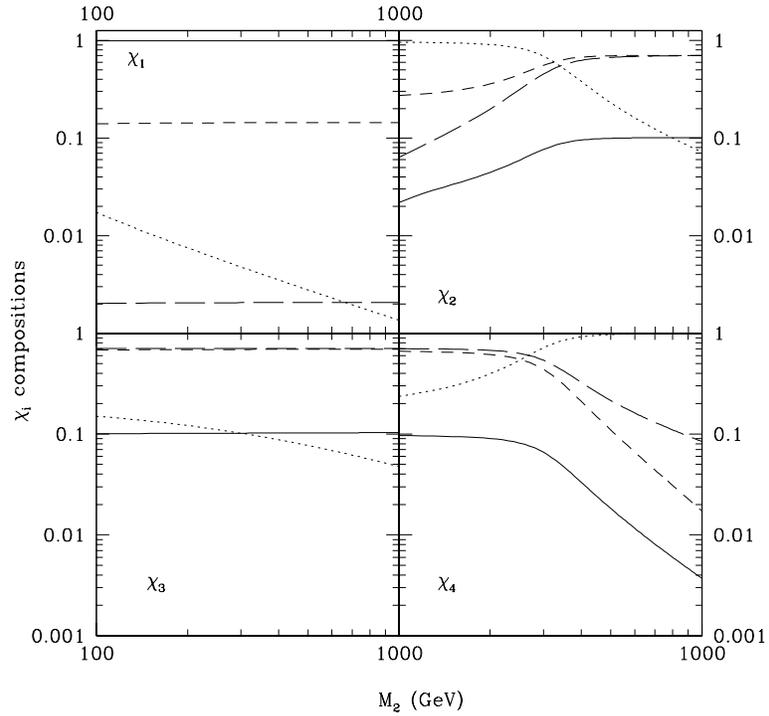}
\end{center}
\caption{Compositions of the  neutralino states $\chi_i$ as functions of $M_2$ for $M_1=25$ GeV,
$\mu=300$ GeV and $\tan\beta=$10. Up--left panel: $\chi_1$,
up--right panel: $\chi_2$, bottom--left panel: $\chi_3$,
bottom--right panel: $\chi_4$.
Solid lines denote $|a_1^{i}|$, dotted lines
$|a_2^{i}|$, short--dashed lines $|a_3^{i}|$, long--dashed lines $|a_4^{i}|$.
\label{fig:neutralino_compositions}.}
\end{figure}

\begin{figure}
\begin{center}
\includegraphics[width=8cm, bb= 80 270 533 533]{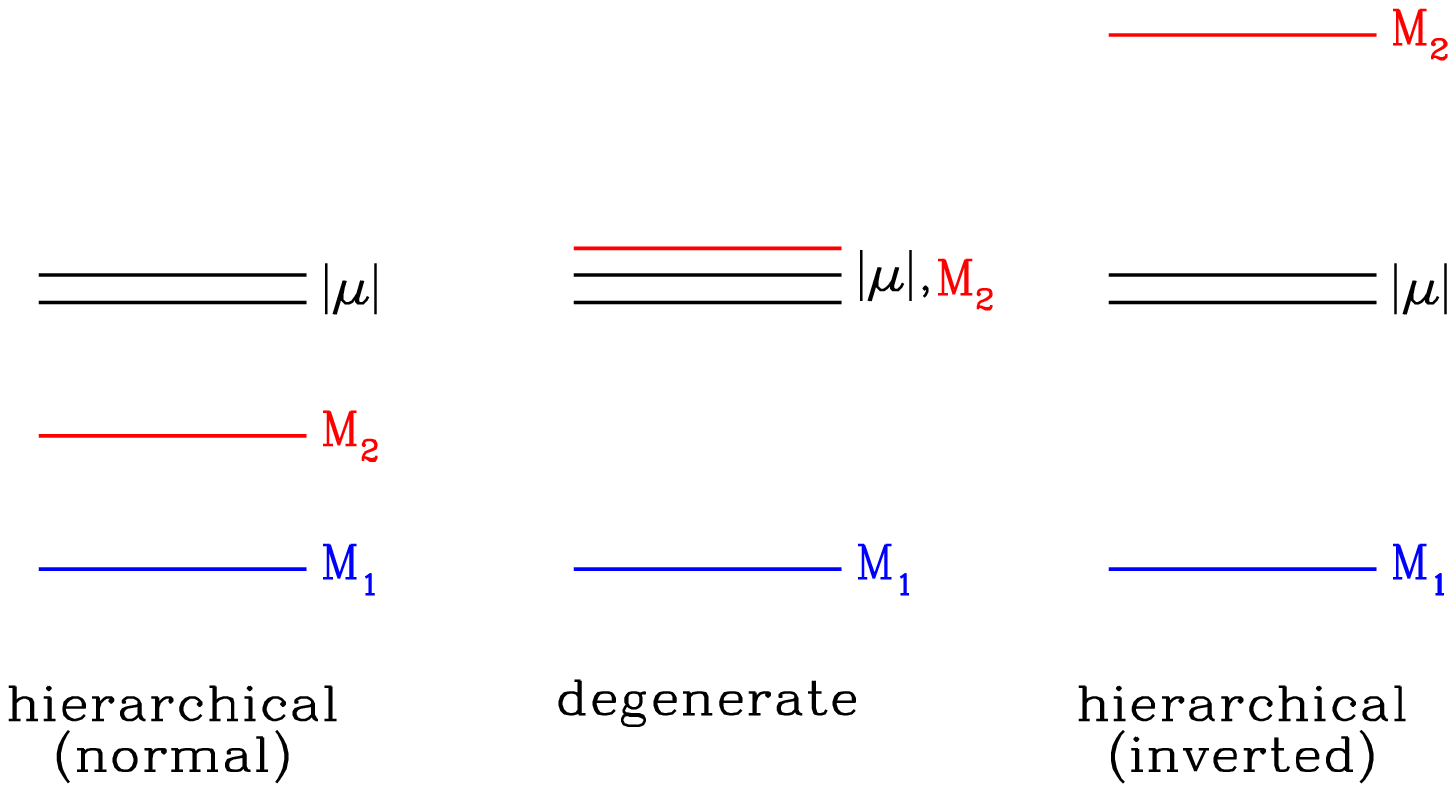}
\end{center}
\caption{
Asymptotic spectroscopic schemes for the $\chi_i$ ($i=1,2,3,4$)
neutralino states
\label{fig:scenarios}.}
\end{figure}

For illustrative purposes, in
Figs.\ref{fig:neutralino_masses}--\ref{fig:neutralino_compositions} we
show the results of the numerical diagonalization of the matrix of
Eq. (\ref{massmatrix}) versus $M_2$, for the representative point
defined by the following values of the other supersymmetric parameters
which enter in the mass matrix of Eq. (\ref{massmatrix}): $M_1$ = 25
GeV, $\mu$ = 300 GeV, $\tan \beta$ = 10.
Fig. \ref{fig:neutralino_masses} displays the behaviour of the mass
eigenvalues $|m_i|$, Fig. \ref{fig:neutralino_compositions} the compositions of the states $\chi_i$.

The approximate analytic expressions in Eqs. (\ref{approx1}-\ref{approx4}) and the numerical results in
Fig. \ref{fig:neutralino_compositions}
 display quantitatively the following properties: (i) $\chi_1$ is mainly a B-ino whose mixing
with $\tilde H_1^{\circ}$ is sizable at small $\mu$, (ii) $\chi_3$ has a mass $|m_3| \simeq |\mu|$ with a
large $\tilde H_1^{\circ}-\tilde H_2^{\circ}$ mixing, independently of $M_2$, (iii) $\chi_2$ and $\chi_4$
interchange their main structures depending on the value of the ratio $|\mu|/M_2$: $\chi_2$ is dominantly a W-ino
(with a sizable subdominance of $\tilde H_1^{\circ}$) for $M_2 << |\mu|$ and a maximal
$\tilde H_1^{\circ}-\tilde H_2^{\circ}$ admixture for $M_2 >> |\mu|$, whereas $\chi_4$ is  a maximal
 $\tilde H_1^{\circ}-\tilde H_2^{\circ}$ admixture for $M_2 << |\mu|$ and a very pure W-ino
 for $M_2 >> |\mu|$. Also their properties relevant to the case $M_2 \sim |\mu|$ are transparent from
 Eqs. (\ref{approx1}-\ref{approx4}) and
 Fig. \ref{fig:neutralino_compositions}.

Depending on the relative values of the parameters $M_2$ and $\mu$, it
is useful, for the discussion to be developed later, to define the
following neutralino spectroscopic schemes (notice that we always
assume ($M_1 << M_2, |\mu|$): (i) normal hierarchical scheme ($M_2 <
|\mu|$), (ii) degenerate scheme ($M_2 \sim |\mu|$), (iii) inverted
hierarchical scheme ($M_2 > |\mu|$). These schemes are depicted in
Fig. \ref{fig:scenarios}.

\subsection{Cosmological properties}
\label{sec:cosmological_properties}

As discussed in Ref. \cite{lowneu}, under the assumption that R-parity is conserved,  the lower
bound on the neutralino mass $m_{\chi}$ is provided by the upper bound on the relic abundance
for cold dark matter (CDM), $\Omega_{CDM} h^2$.

We recall that the neutralino relic abundance is given by

\begin{equation}
\Omega_{\chi} h^2 = \frac{x_f}{{g_{\star}(x_f)}^{1/2}} \frac{3.3 \cdot
10^{-38} \; {\rm cm}^2}{\widetilde{<\sigma_{ann} v>}},
\label{omega}
\end{equation}

\noindent
where $\widetilde{<\sigma_{ann} v>} \equiv x_f {\langle \sigma_{\rm
ann} \; v\rangle_{\rm int}}$, ${\langle \sigma_{\rm ann} \;
v\rangle_{\rm int}}$ being the integral from the present temperature up to
the freeze-out temperature $T_f$ of the thermally averaged product of
the annihilation cross-section times the relative velocity of a pair
of neutralinos, $x_f$ is defined as $x_f \equiv \frac{m_{\chi}}{T_f}$ and
${g_{\star}(x_f)}$ denotes the relativistic degrees of freedom of the
thermodynamic
bath at $x_f$.  For $\widetilde{\langle \sigma_{\rm ann} \; v\rangle}$
we use here the standard expansion in S and P waves:
$\widetilde{\langle \sigma_{\rm ann} \; v\rangle}\simeq \tilde{a} +
 \tilde{b} (2 x_f)^{-1}$.

A host of cosmological observations imply that the cold dark matter content
has to stay in the range
 $0.092 \leq \Omega_{CDM} h^2 \leq 0.124$ \cite{wmapetc}. Thus the
 supersymmetric configurations have to satisfy the
 cosmological constraint
 $\Omega_{\chi} h^2 \leq (\Omega_{CDM} h^2)_{max} = 0.124$.

\section{Scenarios for light neutralinos}

\label{sec:scenarios}
In Ref. \cite{lowneu} it is shown that the cosmological condition
$\Omega_{\chi} h^2 \leq (\Omega_{CDM} h^2)_{max} = 0.124$ provides a
lower limit $m_{\chi} \gsim$ 7 GeV. Such a situation occurs when $M_1
\sim$ 10 GeV and $\widetilde{\langle \sigma_{\rm ann} \;
v\rangle}\simeq \tilde{a}$ receives a sizable contribution by the
exchange of the A Higgs boson in the {s} channel.  This, in turn,
happens when the two following conditions are met: a) $m_A$ is as
small as its experimental lower bound, $m_A = $90 GeV, b) the B-ino
component of the $\chi_1$ configuration is maximally mixed with the
$\tilde H_1^{\circ}$ component ({\it i. e.}
$\frac{a_3^{(1)}}{a_1^{(1)}} \simeq 0.4)$. From the second expression
in Eq. (\ref{approx1}) one sees that condition (b) is satisfied when
$\mu$ is small ($|\mu| \sim$ 100-200 GeV).  Moreover, it turns out
that $\tan \beta$ must be large ($\tan \beta \sim$ 30 - 45) in order
for the annihilation cross section through the exchange of the A Higgs
boson to be sizable. Finally, the trilinear coupling is mildly
constrained to stay in the interval $-1 \lsim A \lsim +1$.

In the following we will characterize such a set of supersymmetric
parameters as {\bf Scenario {$\mathcal{A}$}}. More specifically, this
scenario is identified by the following sector of the supersymmetric
parameter space: $M_1 \sim$ 10 GeV, $|\mu| \sim$ (100 - 200) GeV, $m_A
\sim$ 90 GeV, $\tan \beta \sim$ 30 - 45, -1 $\lsim A \lsim$ +1; the
other supersymmetric parameters are not {\it a priori} fixed.

From Eqs. (\ref{approx1}) it turns out that, in this scenario, the following hierarchy holds for the
coefficients $a^{(1)}_i$  of $\chi_1$

\begin{equation}
|a_1^{(1)}| > |a_3^{(1)}| >> |a_2^{(1)}|, |a_4  ^{(1)}|.
\label{hierarchy1}
\end{equation}

It is proved in Ref. \cite{lowneu} that at small $m{\chi}$, when $m_A \gsim$ (200 - 300) GeV, the cosmological
constraint $\Omega_{\chi} h^2 \leq (\Omega_{CD M} h^2)_{max}$ is satisfied because of
 stau-exchange contributions
(in the {\it t, u} channels) to ${\widetilde{<\sigma_{ann} v>}}$, provided that: (i) $m_{\tilde{\tau}}$ is sufficiently
light, $m_{\tilde{\tau}} \sim$ 90 GeV (notice that the current experimental limit is $m_{\tilde{\tau}} \sim$ 87 GeV) and
(ii) $\chi_1$ is a very pure B-ino ({\it i.e.} $(1 - a^{(1)}_1)$ = O($10^{-3}$).
In such a situation, the lower limit on the neutralino mass is $m{\chi} \gsim$ (15 - 18) Gev \cite{others}.

Let us first discuss the implications of the prerequisite (i). The
experimental lower bounds on the sneutrino mass and on the charged
slepton masses of the first two families imply a lower bound on the
soft slepton mass: $m_{\tilde{l}} \gsim$ 115 GeV. In order to make the
request $m_{\tilde{\tau}} \sim$ 90 GeV compatible with $m_{\tilde{l}}
\gsim$ 115 GeV, it is necessary that the off-diagonal terms of the
sleptonic mass matrix in the eigenstate basis, which are proportional
to $\mu \tan \beta$, are large. Numerically, one finds $|\mu| \tan
\beta \sim$ 5000 GeV.

On the other side, the condition (ii) requires that $a^{(1)}_3/a^{(1)} \lsim 10^{-1}$, {\it i. e.},
according to the second expression of Eq. (\ref{approx1}),
$\frac{a_3^{(1)}}{a_1^{(1)}} \simeq s_\theta s_\beta \frac{m_Z}{\mu} \lsim 10^{-1}$. Combining this last
expression with the condition $|\mu| \tan \beta \sim$ 5000 GeV, one finds that $|\mu|$ and $\tan \beta$ are bounded
by: $|\mu| \gsim$ 500 GeV, $\tan \beta \lsim$ 10. These bounds are somewhat weaker for
 values of the neutralino mass larger than $\sim$ 15--18 GeV.

The previous arguments lead us to introduce a new scenario, denoted as
{\bf Scenario $\mathcal{B}$}, identified by the following sector of the supersymmetric
parameter space: $M_1 \sim$ 25 GeV, $|\mu| \gsim$ 500 GeV, $\tan \beta
\lsim$ 20; $m_{\tilde{l}} \gsim$ (100 - 200) GeV, $-2.5 \lsim A \lsim +2.5$;
the other supersymmetric parameters are not {\it a priori} fixed.

From Eqs. (\ref{approx1}) it turns out that, in this scenario, the
following hierarchy holds for the coefficients $a^{(1)}_i$ of $\chi_1$

\begin{equation}
|a_1^{(1)}| >> |a_3^{(1)}|, |a_2^{(1)}|, |a_4  ^{(1)}|.
\label{hierarchy2}
\end{equation}

The features of scenarios $\mathcal{A}$ and $\mathcal{B}$ are summarized in Table
\ref{table:scenarios}.

\begin{table}[t]
\begin{center}
{\begin{tabular}{@{}|c|c|c|c|c|c|@{}}
\hline
~~{\rm scenario}~~ &  ~~~~$M_1$~[GeV]~~~~  & $|\mu|$~[GeV]  & $\tan\beta$   &  $m_A$~[GeV] &
$m_{\tilde{l}}~[\rm GeV]$
\\
\hline
\hline
{$\mathcal{A}$} &  $\sim$ 10  & 100--200 & 30--45 & $\sim$ 90  & --
\\
{$\mathcal{B}$} &  $\sim$ 25  & $\gsim$ 500 & $\lsim$ 20  & $\gsim$
200   & 100--200 \\
\hline
\end{tabular}}
\caption{Scenarios for light neutralinos as described in Section
\protect\ref{sec:scenarios}. In scenario $\mathcal{A}$ the slepton soft mass $m_{\tilde{l}}$
is unconstrained: in our analysis a few representative values for
$m_{\tilde{l}}$ are considered within its natural range 115 GeV $\leq
m_{\tilde{l}}\leq$ 1 TeV. Moreover, in scenario $\mathcal{A}$: -1 $\lsim A
\lsim$ +1,
in scenario $\mathcal{B}$: -2 $\lsim A \lsim$ +2.
\label{table:scenarios}}
\end{center}
\end{table}

\section{Signals at LHC}
\label{sec:signals}

If kinematically accessible, squarks and gluinos are expected to be
copiously produced in the $pp$ scattering process at the LHC.  In the
present paper we limit our discussion to the case in which the gluino
is heavier than the squark; for definiteness, we set
the SU(3) gaugino mass at the representative value $M_3$ = 2 TeV and
the squark soft-mass at the
 value $m_{\tilde{q}}$ = 1 TeV.  Notice that these two parameters
are irrelevant in the specification of the scenarios previously defined.

Many experimental studies are available
showing that the production of 1~TeV squarks can be easily
discovered at the LHC through inclusive analyses based on the request 
of a high multiplicity of hard jets ant $E_T^{miss}$, see 
e.g \cite{atltdr} and \cite{cmstdr2}.\par
  In order to check the supersymmetric model described in the previous 
 sections, 
the  SUSY parameters need to be measured. For this purpose, we employ here  
a strategy which has been developed for the measurement of the
masses of the SUSY particles based on the identification of 
exclusive decay chains consisting in sequences of two-body
decays  \cite{Bachacou:2000zb,atltdr,Allanach:2000kt,Gjelsten:2004ki} . 
The most promising decay chains considered
in the literature are:

\begin{equation}
\tilde{q}\rightarrow q \chi_{i}\rightarrow q \tilde{f}f\rightarrow q
\bar{f}f \chi_1 \;\;\;{\rm (sequential\;\; chain)},
\label{eq:sequential}
\end{equation}

\noindent
and

\begin{equation}
\tilde{q}\rightarrow q \chi_{i}\rightarrow q (Z,h,H,A) \chi_1\rightarrow q
\bar{f}f \chi_1 \;\;\;{\rm (branched\;\; chain)},
\label{eq:branched}
\end{equation}

\noindent
where $f$ stands for a fermion, and the neutralino
subscript $i$ can take the values 2, 3 or 4. 
 From the experimental point of 
view, given the large multiplicity of QCD jets present in $pp$ 
events, the only interesting decays which can be used for the
identification of exclusive chains are the ones involving light 
charged leptons ($e$ and $\mu$), the hadronic decays of $\tau$
($\tau$-jets) and the fragmentation of $b$ quarks ($b$-jets), 
which can be experimentally separated from the background.
The sequential chain
with two $e$ or $\mu$ in the final state is particularly useful, 
as the whole chain consists of three successive 2-body 
decays which can be measured very well, and provides 
enough constraints to allow a model-independent mass determination.
The sequential chain with two $\tau$-jets allows a much
less clean measurement, as neutrinos are produced in the decay 
of the $\tau$ which make the event kinematic less clear, and
a much higher jet background is present. It is however 
important to detect this chain as well, because by comparing
its rate with the rate for the $e$, $\mu$ chain, information
can be extracted on the $\tilde{\tau}$ mixing \cite{Nojiri:2005ph}.
Finally, the branched chain provides less constraints on the
SUSY masses, as the invariant mass of the two final-state
fermions just shows a peak at the value of the resonance
appearing in the decay. In this case the experimentally interesting
decays are the ones in $b$ pairs, especially for the Higgs bosons,
and in $e$ and $\mu$ for the $Z$.\par

These two decay chains have in common the first step consisting in the
squark decay $\tilde{q}\rightarrow q \chi_i$. Two types of
couplings occur in this process: (a) gauge couplings, universal in the
quark flavor, and (b) Yukawa couplings, hierarchical in the quark
mass. Gauge (Yukawa) couplings are proportional to the gaugino
(Higgsino) components in $\chi_i$, respectively.  The actual
number of the $q \chi_i$ states produced in a pp collision depends
on the product of the cross section for the production process
$\sigma(pp\rightarrow
\tilde{q}\tilde{q},\tilde{q}\tilde{q}^*,\tilde{g}\tilde{g},
\tilde{q}\tilde{g})$ times the branching ratio
BR($\tilde{q}\rightarrow q \chi_i$).  In this product the
contribution of the heavy quarks is strongly suppressed, because of
their scarcity in the proton composition. Moreover, since the relative
importance of the Yukawa couplings as compared to the gauge couplings
depends on the ratio $m_q/m_Z$ ($m_q$ and $m_Z$ being the quark mass
and the Z-boson mass, respectively) the production of $\chi_i$ is
more sizable when gauge couplings are effective, {i. e.}  when
$\chi_i$ has large gaugino components.  To account for these
properties it is convenient to define an {\it effective} branching
ratio BR($\tilde{q}\rightarrow q \chi_i$), {\it i. e.}  an average
of the branching rations BR($\tilde{q}\rightarrow q \chi_i$)'s
over the light quarks \cite{cinque}, including both
$\tilde{q}_R$ and $\tilde{q}_L$.  To be conservative, we will use
as an effective branching ratio BR($\tilde{q}\rightarrow q \chi_i$)
the average over the four lightest quarks. From now on we will simply
denote this average as BR($\tilde{q}\rightarrow q \chi_i$).

We turn now to a discussion of the decay process for $\chi_i$,
which takes different routes in sequential and branched chains,
respectively.  In the sequential case, since we have taken
$m_{\tilde{q}}$ = 1 TeV, the decay can only proceed through a slepton:
$\chi_i\rightarrow \tilde{l}l\rightarrow \bar{l}l \chi_1$ with a
branching ratio BR($\chi_i\rightarrow \tilde{l}l\rightarrow
\bar{l}l \chi_1$) = BR($\chi_i\rightarrow \tilde{l}l$)
BR($\tilde{l}\rightarrow l \chi_1$).  The size of
BR($\chi_i\rightarrow \tilde{l}l$) depends sensitively on the
$\chi_i$ composition.  If $\chi_i$ is dominantly a gaugino,
because of the universality of the gaugino couplings, the branching
ratios BR($\chi_i\rightarrow \tilde{l}l$) for the three lepton
flavours are about the same ; if $\chi_i$ is dominantly a
Higgsino, $\chi_i$ decays predominantly into a  
$\tilde{\tau} \tau$ pair.  In the branched chain $\chi_i$ decays either
through the Z-boson or through a Higgs boson. The first case, {\it
i. e.}  $\chi_i \rightarrow Z + \chi_1$, involves only the
Higgsino components of the two neutralino states; the 
$Z$ boson subsequently decays 
into all (kinematically
possible) $\bar{f} f$ pairs according to the Standard Model
branching fractions. The second case, {\it i. e.}
$\chi_i \rightarrow (h,A,H) + \chi_1$, in order to have a sizable
BR, requires that one neutralino state is dominantly a gaugino, the
other dominantly a Higgsino.  Since in our scenarios $\chi_1$ is
dominantly a B-ino state, $\chi_i \rightarrow (h,A,H) + \chi_1$ is
of interest when $\chi_i$ is dominated by the Higgsino
components. Because of the hierarchical character of the Yukawa
coupling, the subsequent decays of the Higgs bosons are dominated by
the production of a $b$ -- $\bar{b}$ pair.  We note that in the case
of a pronounced hierarchical (inverted) scheme (see
Fig. \ref{fig:scenarios}) the direct $\chi_4\rightarrow (h,H,A,Z)
\chi_1$ decay is suppressed because both $\chi_4$ and $\chi_1$ are
dominantly gauginos. This entails that the ``long'' chains
$\chi_4\rightarrow (h,H,A,Z) \chi_{2,3}\rightarrow (h,H,A,Z) \chi_{1}$
can give a sizable contribution. Indeed,   
for both the sequential and branched decays, longer multi-step
decay chains, of the type {\it e.g.}  
$\chi_4\rightarrow X \chi_{2,3}  \rightarrow X X^\prime \chi_1$, 
are in principle very interesting as they provide additional
kinematic constraints. In practice, the superposition of 
many decays with the same final state may be extremely difficult
to disentangle experimentally, and will thus add confusion 
rather than information. For this study we will therefore
limit ourselves to studying the branching ration for direct decays.\par

 In the following sections, for the sequential 
decays we will consider the two cases $f=e,\tau$. We will first
analyze $e$, for which the two chirality states $\tilde{e}_R$ and $\tilde{e}_L$
will be considered separately. The decay $\chi_i \rightarrow e \tilde{e}$
is largely dominated by the gaugino components in $\chi_i$; thus,  
if the $\chi_i$ decays into both $\tilde{e_R}$ and
$\tilde{e}_L$ are kinematically allowed, the ratio of the two
branching ratios scales approximately as 
$BR(\chi_i\rightarrow e \tilde{e}_L)/BR(\chi_i\rightarrow e
\tilde{e}_R)\simeq 1/4 (1+cot_{\theta} a_2^{(i)}/a_1^{(i)})^2$, where
$a_2^{(i)}/a_1^{(i)}$ can be approximated by
Eqs. (\ref{approx23},\ref{approx4}). This implies that $\chi_i$ decays
dominantly to $\tilde{e_L}$ when it is a pure wino
($a_2^{(i)}/a_1^{(i)}\gg 1$), while either of the two selectron final
states $\tilde{e}_{L,R}$ may be important for other compositions. For the 
$\tau$ case we will 
consider both the mixed (eigenmass) states $\tilde{\tau}_1$ and $\tilde{\tau}_2$.
The presence of both states in the decay can be hard to disentangle,
and thus make the mass and rate measurements more difficult.
For the branched decays we consider the decay into $b$ pairs for 
the Higgs bosons and the decays into $b$ and $e$ pairs for the $Z$.
 In most of the considered models the $Z$ and the SUSY Higgses are almost 
degenerate, and, given the experimental resolution on the $b\bar{b}$ peak, 
 cannot be separated. It is therefore very useful to have 
also the decay into leptons, which will allow the experiments 
to determine the presence of SUSY Higgses in the decay chains.

Based on this discussion, we report our numerical results for all the decay branching ratios
relevant for the decay chains of Eqs. (\ref{eq:sequential}- \ref{eq:branched})
in a number of benchmarks within the scenarios $\mathcal{A}$ and $\mathcal{B}$, previously
defined (see Table \ref{table:benchmarks}).
In these benchmarks, all values of the supersymmetric parameters, except $M_2$, are fixed.
The branching ratios, given as functions of $M_2$, have been calculated
by using the ISASUSY code\cite{isasusy},

\subsection{Sequential chain benchmarks}
\label{sec:sequential}

In scenario $\mathcal{A}$ the parameter $|\mu|$  is required
to be $\mu \sim$ (100-200) GeV, while slepton masses are unconstrained.  On the other
hand the sequential chain is sensitive to the hierarchy between
$|\mu|$ and $m_{\tilde{l}}$, since, when $m_{\tilde{l}} > |\mu|$, the
 decay of the $\chi_{2,3}$ states is not allowed by
kinematics, and the process (\ref{eq:sequential}) can proceed only
through $\chi_4$. On the contrary, when $ m_{\tilde{l}} < |\mu|$, the
three states $\chi_i$ ($i=2,3,4$) take part in the chain
(\ref{eq:sequential}).

\begin{figure}
\includegraphics[width=5.5cm,bb= 0 340 590 540,clip=true]{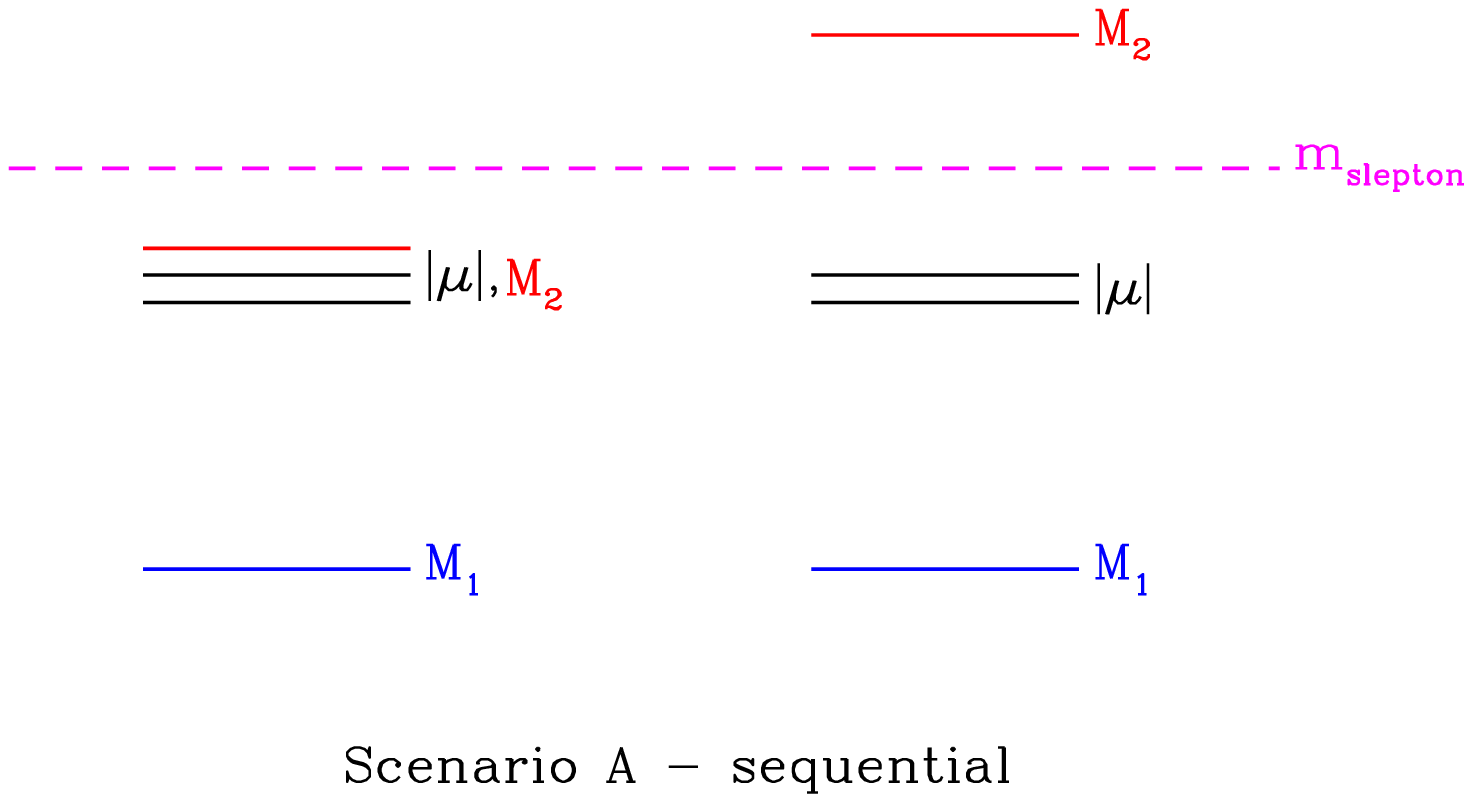}
\includegraphics[width=5.5cm,bb= 0 340 590 540,clip=true]{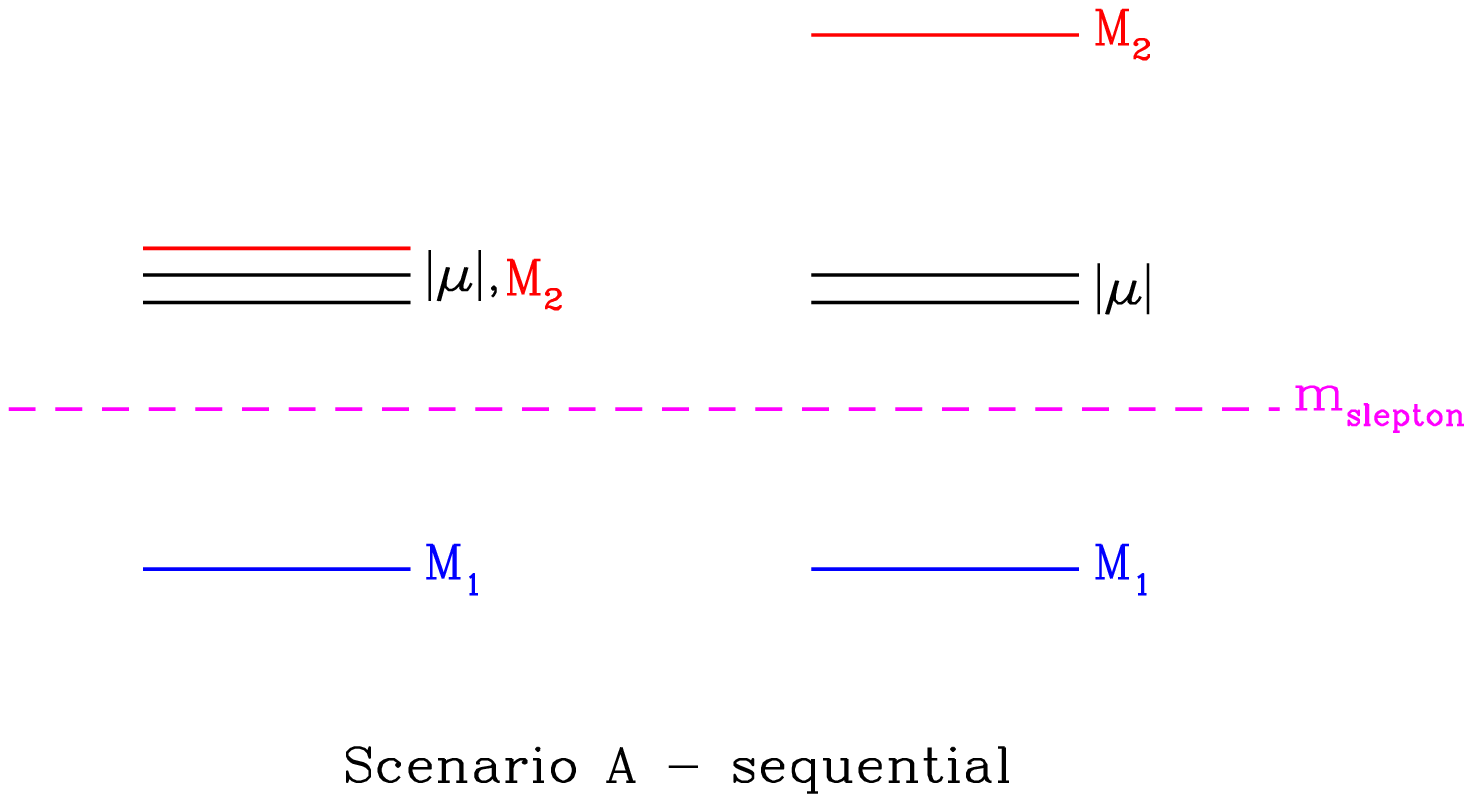}
\includegraphics[width=5.5cm,bb= 0 340 590 540,clip=true]{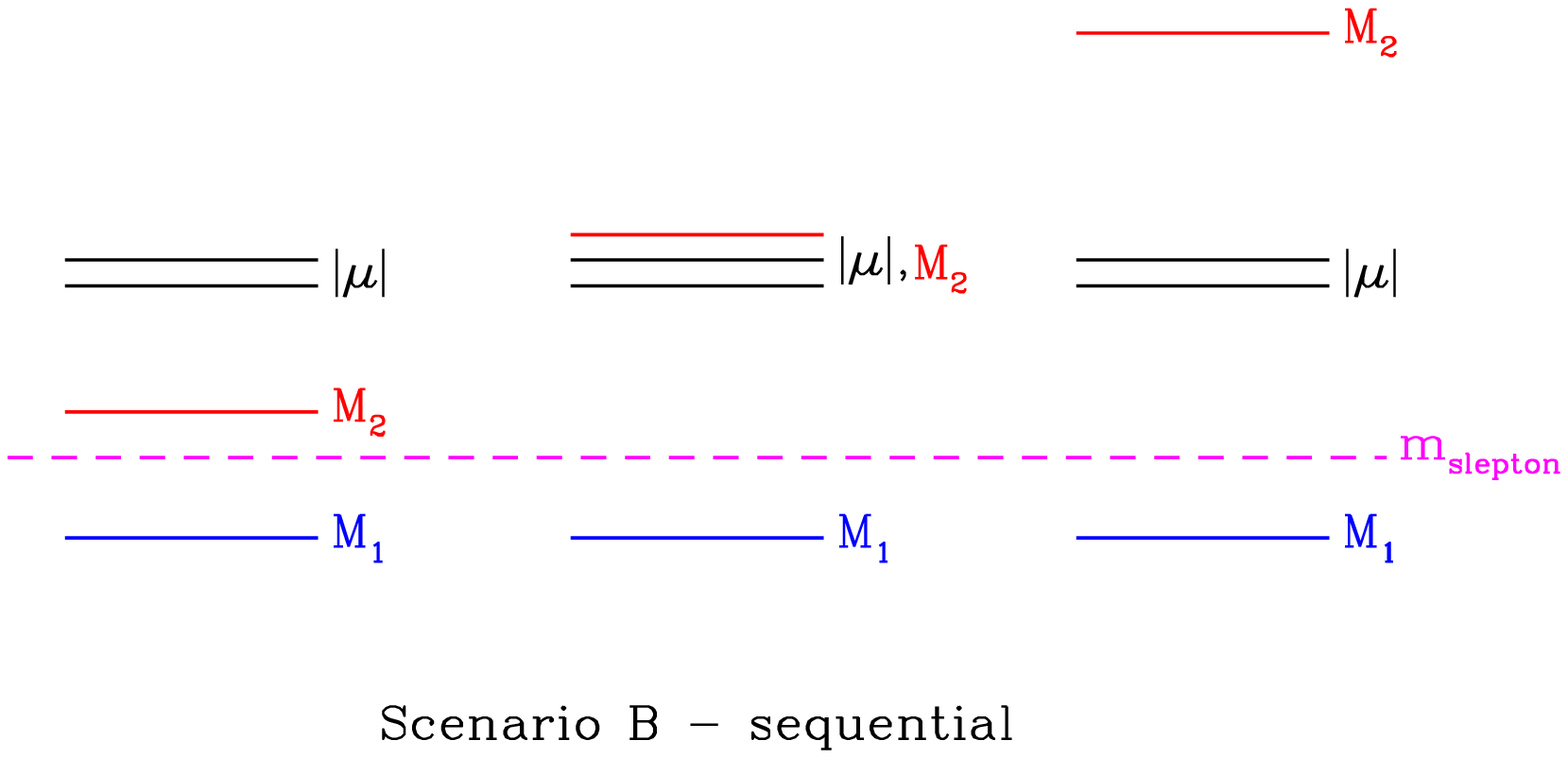}
\caption{Qualitative schemes for benchmarks in sequential decay
chains: $\mathcal{A}$--seq1, $\mathcal{A}$--seq2, $\mathcal{B}$--seq. For each benchmark, extremes values for $M_2$
are considered: $M_2 \sim |\mu|$ and $M_2 > |\mu|$ for scenario $\mathcal{A}$;
$M_2 < |\mu|$, $M_2 \sim |\mu|$ and $M_2 > |\mu|$ for scenario $\mathcal{B}$.
\label{fig:sequential}.}
\end{figure}

We then fix the following benchmarks for scenario $\mathcal{A}$:

\begin{eqnarray}
{\bf \mathcal{A}{\rm -seq1:}} & \mu=110 \; {\rm GeV}  & m_{\tilde{l}}={\rm 150,
  300, 500, 700 \; GeV}
\label{eq:A-seq1}\\
{\bf \mathcal{A}{\rm -seq2:}} & \mu=150 \; {\rm GeV} & m_{\tilde{l}}=120 \;{\rm GeV}, \label{eq:A-seq2}
\end{eqnarray}
\noindent where, in both cases, $M_1$=10 GeV, $\tan\beta$=35, $m_A=$90 GeV
and $A$=0.  These two benchmarks are depicted qualitatively in
Fig.\ref{fig:sequential} and summarized in Table \ref{table:benchmarks}.

\begin{table}[t]
\begin{center}
{\begin{tabular}{@{}|c|c|c|c|c|c|@{}}
\hline
~~{\rm benchmark}~~ &  ~~~~$M_1$~[GeV]~~~~  & $\mu$~[GeV]  & $\tan\beta$   &  $m_A$~[GeV] &
$m_{\tilde{l}}~[\rm GeV]$ \\
\hline
\hline
{\rm $\mathcal{A}$--seq1} &   10  &   110 & 35  & 90   & 150,300,500,700\\
{\rm $\mathcal{A}$--seq2} &   10  &   150 & 35  & 90   & 120  \\
{\rm $\mathcal{A}$--brc}  &   10  &   110 & 35  & 90   & 150  \\
{\rm $\mathcal{B}$--seq}  &   25  &  -500 & 10  & 1000 & 120  \\
{\rm $\mathcal{B}$--brc1} &   25  &  -500 & 10  & 200  & 120  \\
{\rm $\mathcal{B}$--brc2} &   25  &  -500 & 10  & 1000 & 120  \\
\hline
\end{tabular}}
\caption{Benchmarks for light neutralinos,
singled out within the two
scenarios $\mathcal{A}$ and $\mathcal{B}$ of Table \protect\ref{table:scenarios}.
All results will be presented as functions of $M_2$,  $M_2$ being 
 varied in the range
$100$ GeV$\le M_2 \le 1000$ GeV. The other supersymmetric parameters are set at
the representative values: $M_3$ = 2 TeV, $m_{\tilde{q}}$ = 1 TeV, $A$ = 0.
\label{table:benchmarks}}
\end{center}
\end{table}

\begin{figure}
\begin{center}
\includegraphics[width=12cm]{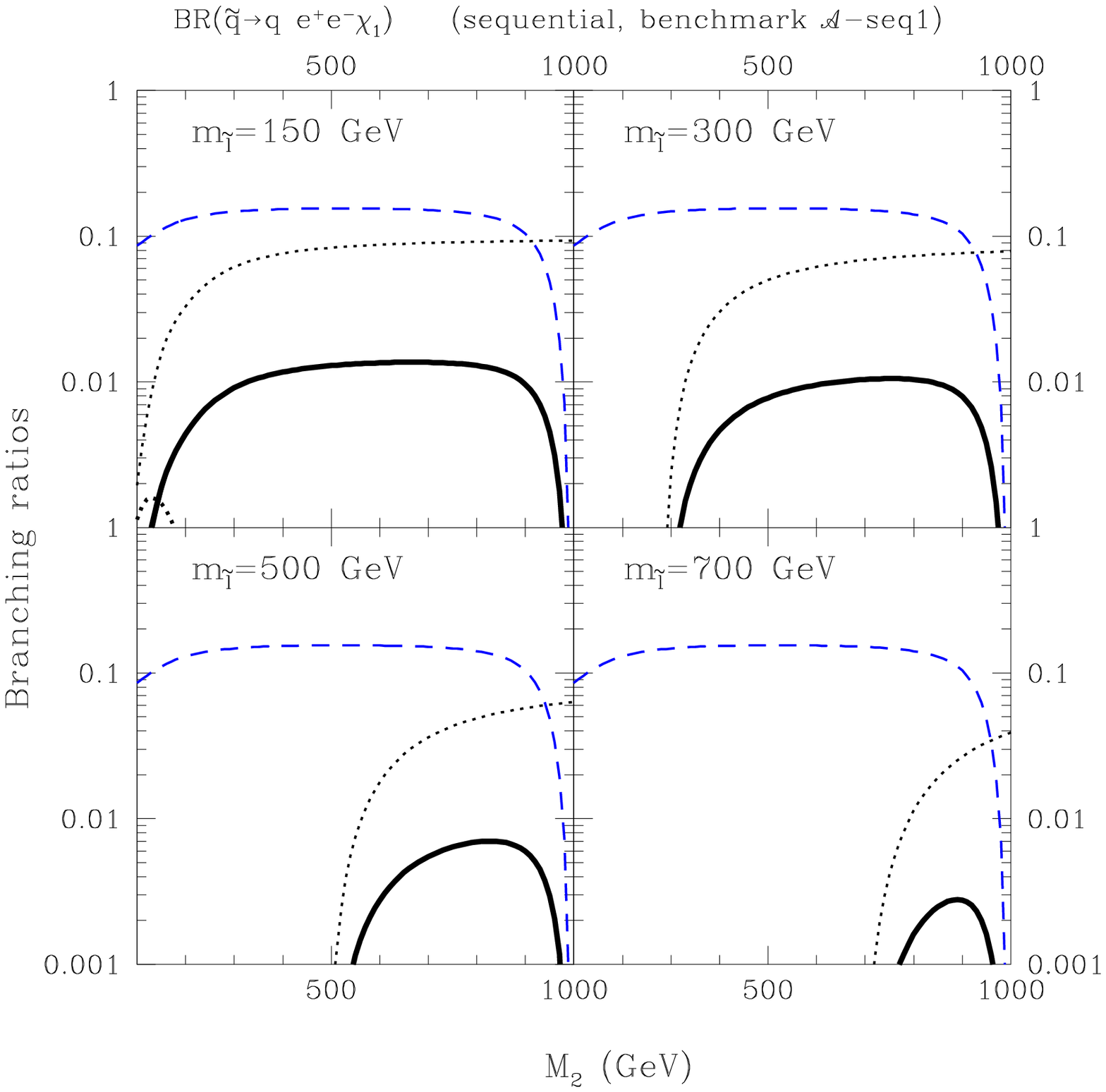}
\end{center}
\caption{Branching ratios for the sequential process $\tilde{q}
\rightarrow \bar{e} e \chi_1$ in benchmark $\mathcal{A}$--seq1 as
functions of $M_2$. Each panel corresponds to a different value of
$m_{\tilde{l}}$.  The dashed lines show the branching ratio for the
process $\tilde{q} \rightarrow q \chi_{4}$; the thin--dotted lines
denote the branching ratio for the process $\chi_{4}\rightarrow e
\tilde{e}_L\rightarrow e\bar{e}\chi_1$; the thick--dotted line in the
top--left panel denotes the branching ratio for the process
$\chi_{4}\rightarrow e \tilde{e}_R\rightarrow e\bar{e}\chi_1$ (the
corresponding curves cannot be seen in the other panels because they are
too low); the thick solid lines denote the branching ratio for the
whole sequential decay chain $\tilde{q}\rightarrow q
\chi_{4}\rightarrow q e \tilde{e}\rightarrow q e \bar{e}\chi_1$.}
\label{fig:aseq1_e}
\end{figure}

\begin{figure}
\begin{center}
\includegraphics[width=12cm]{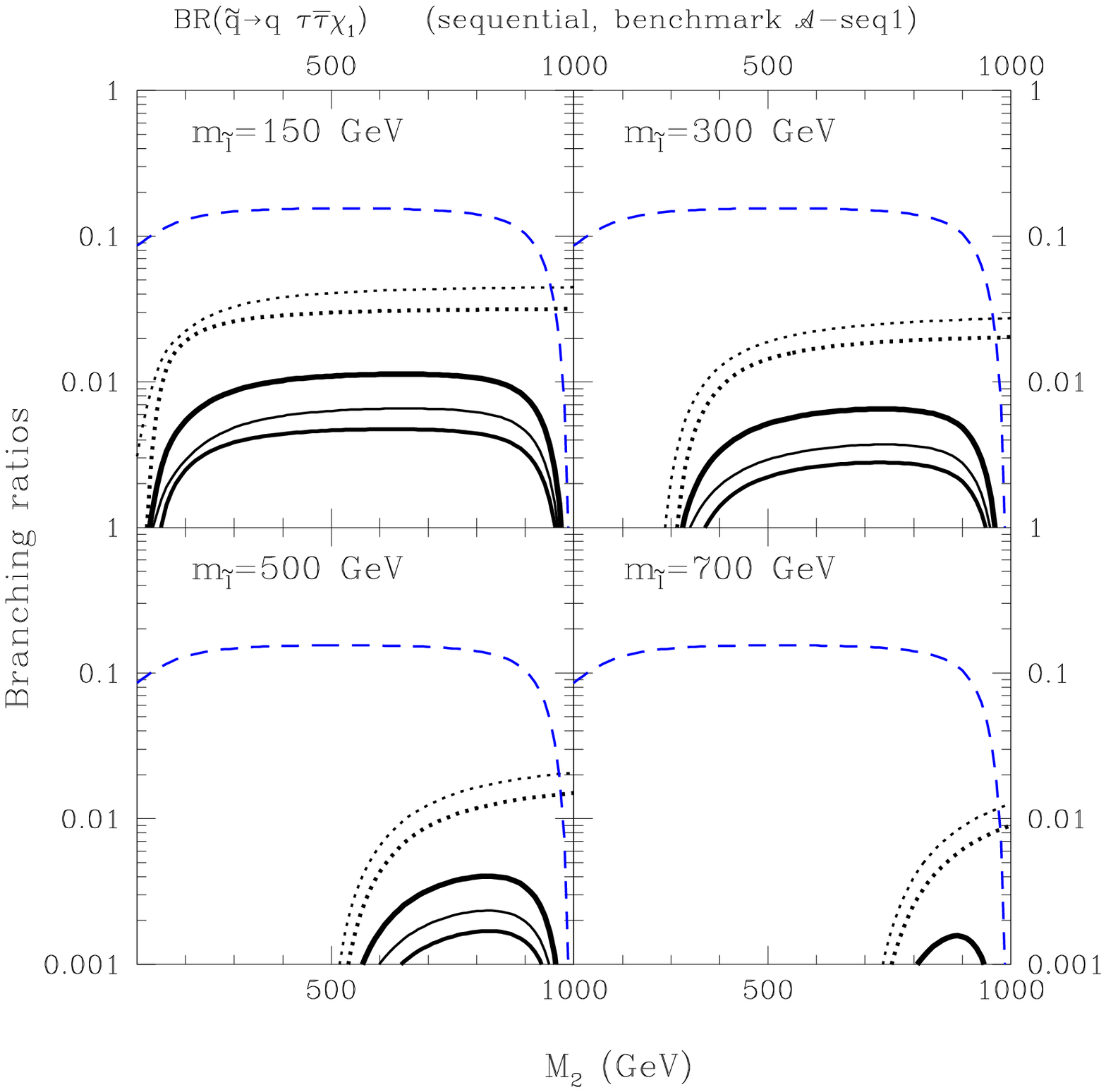}
\end{center}
\caption{Branching ratios for the sequential process $\tilde{q}
\rightarrow \bar{\tau} \tau \chi_1$ in benchmark $\mathcal{A}$--seq1
as functions of $M_2$. Each panel corresponds to a different value of
$m_{\tilde{l}}$.  The dashed lines show the branching ratio for the
process $\tilde{q} \rightarrow q \chi_{4}$; the thin--dotted lines
denote the branching ratio for the process 
$\chi_{4}\rightarrow \tau\tilde{\tau_1} \rightarrow \tau \bar{\tau}\chi_1$; 
the thick--dotted lines denote the branching ratio for the process 
$\chi_{4}\rightarrow \tau\tilde{\tau_2} \rightarrow \tau \bar{\tau}\chi_1$;
the thin--solid lines denote the branching ratio for the whole
sequential decay chain $\tilde{q}\rightarrow q \chi_{4}\rightarrow q
\tau \tilde{\tau_1}\rightarrow q \tau \bar{\tau}\chi_1$; the solid
lines with intermediate thickness denote the branching ratio for the
whole sequential decay chain $\tilde{q}\rightarrow q
\chi_{4}\rightarrow q \tau \tilde{\tau_2}\rightarrow q \tau
\bar{\tau}\chi_1$; the thickest solid lines denote the total branching
ratio for the whole sequential decay chain $\tilde{q}\rightarrow q
\chi_{4}\rightarrow q \tau \tilde{\tau}\rightarrow q \tau
\bar{\tau}\chi_1$.
\label{fig:aseq1_tau}}
\end{figure}

For the benchmark $\mathcal{A}$--seq1 the branching ratios for the
sequential decay chain are shown in Figs.\ref{fig:aseq1_e},\ref{fig:aseq1_tau} where, as
already mentioned, only the decay of $\chi_4$ is kinematically
allowed. Each panel of the figure corresponds to a different value of
$m_{\tilde{l}}$ among those of Eq.(\ref{eq:A-seq1}):
$m_{\tilde{l}}=$150, 300,500,700 GeV. The notations for the various
curves are explained in the figure caption.

The main features of Figs.\ref{fig:aseq1_e}  and 
\ref{fig:aseq1_tau} are readily understood. The Branching Ratio
(BR) for $\tilde{q}\rightarrow q\chi_4$ is $\sim$ 15\%, since the $\chi_4$ 
is dominantly wino, and therefore the $q_R$ has no BR into it,
and the $\tilde{q}_L$ decays 60\% into a chargino  $\chi^\pm_2$ and 30\% into $\chi_4$,
according to the left-handed couplings of the wino.
In the regime $M_2 >> m_{\tilde{l}}, $ the $\chi_4$ decays with 40\% BR into $W$, $Z$ and Higgses
because of its non-zero Higgsino component, 60\% BR into the 
left-handed sleptons.  So for $\tilde{e}_L$ the BR is $\sim$10\%.
For each of the two $\tilde{\tau}$ states, which are in this case
rather similar in composition and mass, the BR is $\sim5\%$ for 
each state.  For instance, 
 for $m_{\tilde{l}}$ = 150 GeV: $m_{\tilde{\tau}_1}$ = 133 GeV, 
$m_{\tilde{\tau}_2}$ = 176 GeV   and  
for $m_{\tilde{l}}$ = 300 GeV: $m_{\tilde{\tau}_1}$ = 192 GeV, 
$m_{\tilde{\tau}_2}$ = 314 GeV.

The different quadrants of Figs. \ref{fig:aseq1_e} and
\ref{fig:aseq1_tau} show by how much the branching ratios are suppressed  
for growing values of $m_{\tilde{l}}$, due to the reduction
of the available phase space for the decay
$\chi_4 \rightarrow l \tilde{l}$. On the other hand, a growing value of
$m_{\tilde{l}}$ does not affect $BR(\tilde{l} \rightarrow l
\chi_1)$, which is very close to 1 in all cases. Note that in
this benchmark whenever $M_2 \gsim |\mu|$ one has $m_{\chi_4}\sim M_2$,
thus a measurement of $m_{\chi_4}$ would typically give direct access to
the value of $M_2$.

\begin{figure}
\begin{center}
\includegraphics[width=12cm]{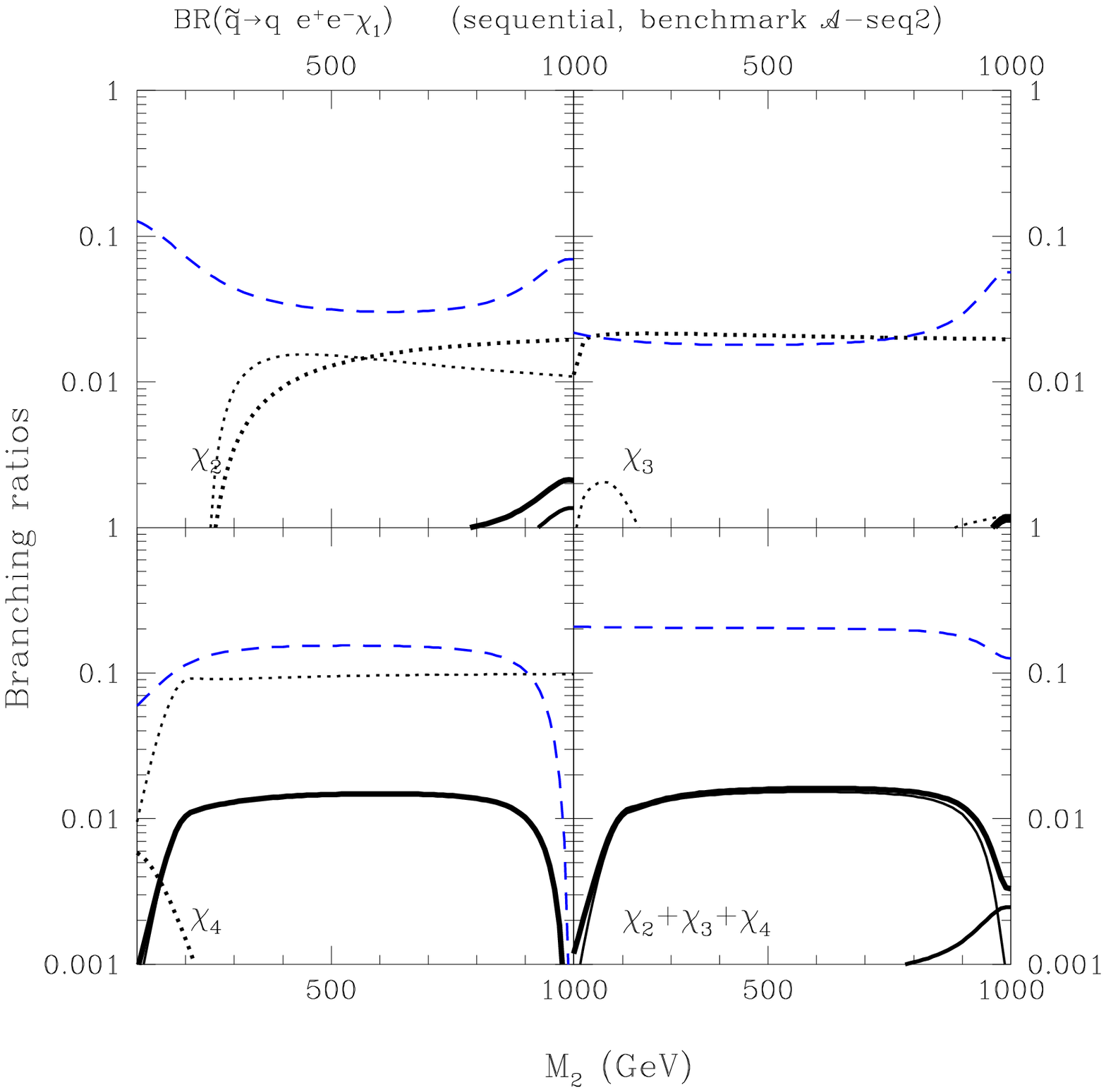}
\end{center}
\caption{Branching ratios for the sequential process
$\tilde{q} \rightarrow \bar{e} e  \chi_1$
in benchmark $\mathcal{A}$--seq2
as functions of $M_2$. Each of the first three panels corresponds to a
different intermediate neutralino state ($\chi_i, i=2,3,4$)
in the sequential chain, while the last panel refers to
combined branching ratios.
The dashed lines show the branching ratio for the
process $\tilde{q} \rightarrow q \chi_{4}$; the thin--dotted lines
denote the branching ratio for the process $\chi_{4}\rightarrow e
\tilde{e}_L\rightarrow e\bar{e}\chi_1$; the thick--dotted lines  
denote the branching ratio for the process
$\chi_{4}\rightarrow e \tilde{e}_R\rightarrow e\bar{e}\chi_1$ 
 The thin--solid lines denote the branching ratio for the
whole
sequential decay chain $\tilde{q}\rightarrow q \chi_{4}\rightarrow q
e \tilde{e}_L\rightarrow q e \bar{e}\chi_1$; the solid--lines 
with intermediate thickness denote the branching ratio for the
whole sequential decay chain $\tilde{q}\rightarrow q
\chi_{4}\rightarrow q e \tilde{e}_R\rightarrow q e
\bar{e}\chi_1$; the thickest solid lines denote the total branching
ratio for the whole sequential decay chain $\tilde{q}\rightarrow q
\chi_{4}\rightarrow q e \tilde{e}\rightarrow q e
\bar{e}\chi_1$ (some curves do not appear in all panels because they
are too low).
\label{fig:aseq2_e}}
\end{figure}

\begin{figure}
\begin{center}
\includegraphics[width=12cm]{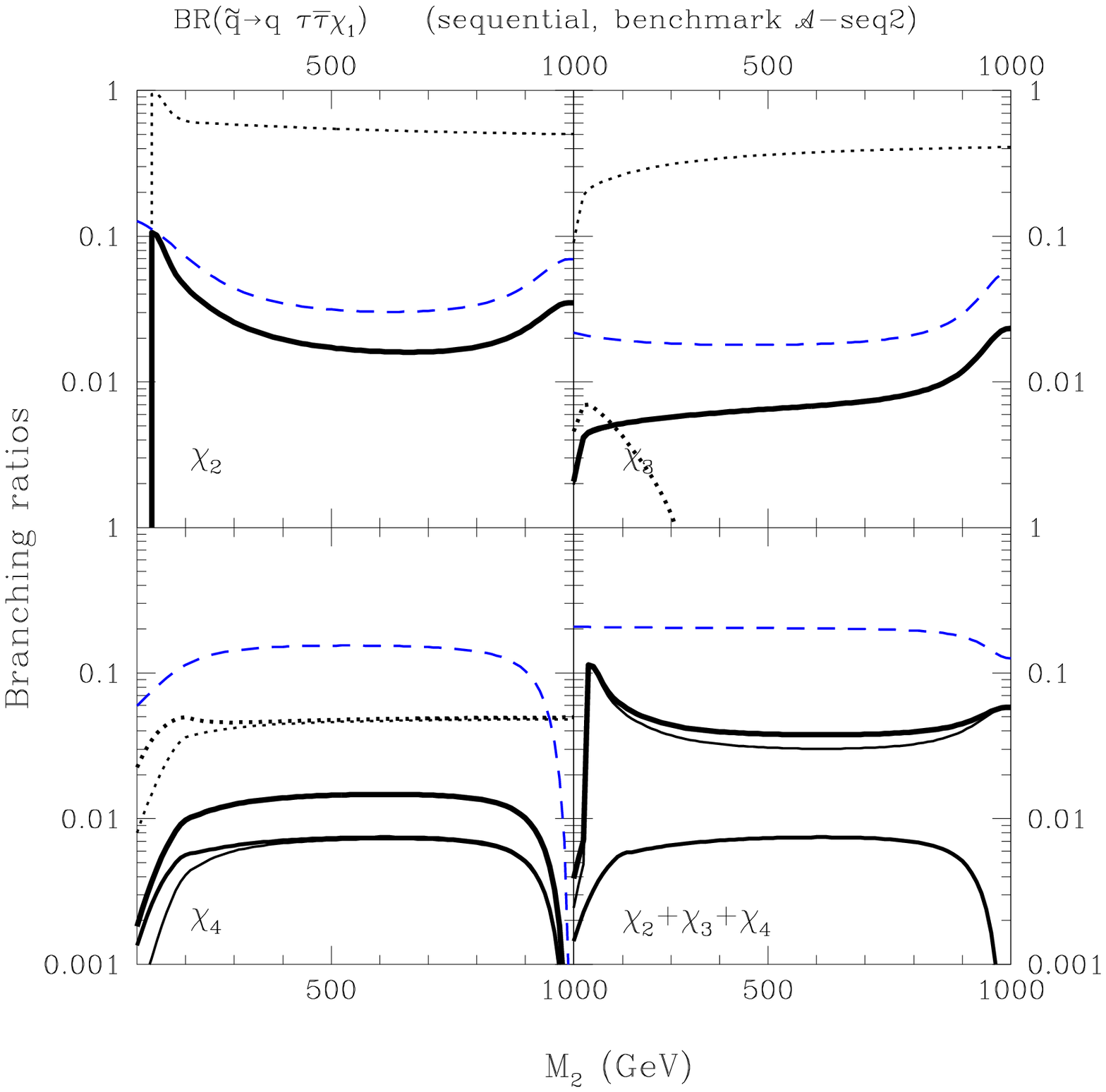}
\end{center}
\caption{Branching ratios for the sequential process  
$\tilde{q} \rightarrow \bar{\tau} \tau  \chi_1$
in benchmark $\mathcal{A}$--seq2
as functions of $M_2$. Each of the first three panels corresponds to a
different intermediate neutralino state ($\chi_i, i=2,3,4$)
in the sequential chain, while the last panel refers to
combined branching ratios.
The codes for the various curves are as in Fig. \ref{fig:aseq1_tau}. 
\label{fig:aseq2_tau}}
\end{figure}

The phenomenology of the sequential decay becomes
richer in the benchmark $\mathcal{A}$--seq2, which is shown in 
Figs.\ref{fig:aseq2_e} and \ref{fig:aseq2_tau}.
In fact in this case the process (\ref{eq:sequential}) can proceed
through the production and decay of any of the $\chi_i$ states. The
corresponding separate contributions to the decay branching ratios for
$i=2,3,4$ are shown in each of the first three panels of 
Figs.\ref{fig:aseq2_e} and \ref{fig:aseq2_tau}
while the last panel shows the combined total branching ratios.

 The branching ratios for the process $\tilde{q} \rightarrow q \chi_{4}$, shown
in the bottom-left panel of 
Figs.\ref{fig:aseq2_e}--\ref{fig:aseq2_tau},
 are obviously very similar the ones already displayed
in Figs.\ref{fig:aseq1_e}--\ref{fig:aseq1_tau}, and previously commented. 
As compared to these,
the branching ratios for the processes  
$\tilde{q} \rightarrow q \chi_{i}$ $(i=2,3)$
are somewhat suppressed, due to the prevalent Higgsino compositions of
$\chi_2, \chi_3$. For the same reason, 
the BR  of $\chi_2$ and $\chi_3$ into $e_{L,R}$, is at most a couple
of percent. The contrary happens for the branching ratios
BR($\chi_i \rightarrow \tilde{\tau} \tau \chi_1$). In this
case $\chi_2, \chi_3$ decay into $\tilde{\tau}\tau$ with larger BRs 
than $\chi_4$, because the dominance 
of the Higgsino components in $\chi_2, \chi_3$
favors their decays into third generation leptons. 
Notice that in the case of  $\chi_2$, only the lighter stau contributes, 
since $\tilde{\tau}_2$ is heavier than  $\chi_2$; in the case of $\chi_3$, 
the $\tilde{\tau}_2$ contribution is very strongly suppressed as compared to 
the $\tilde{\tau}_1$ contribution again for phase--space reasons. 
All in all, in the full process
$\tilde{q}\rightarrow q \chi_{i}\rightarrow q \tilde{\tau}\rightarrow
q \tau \bar{\tau}\chi_1$, the most relevant contribution turns out to be
provided by $\chi_2$.

\begin{figure}
\begin{center}
\includegraphics[width=12cm]{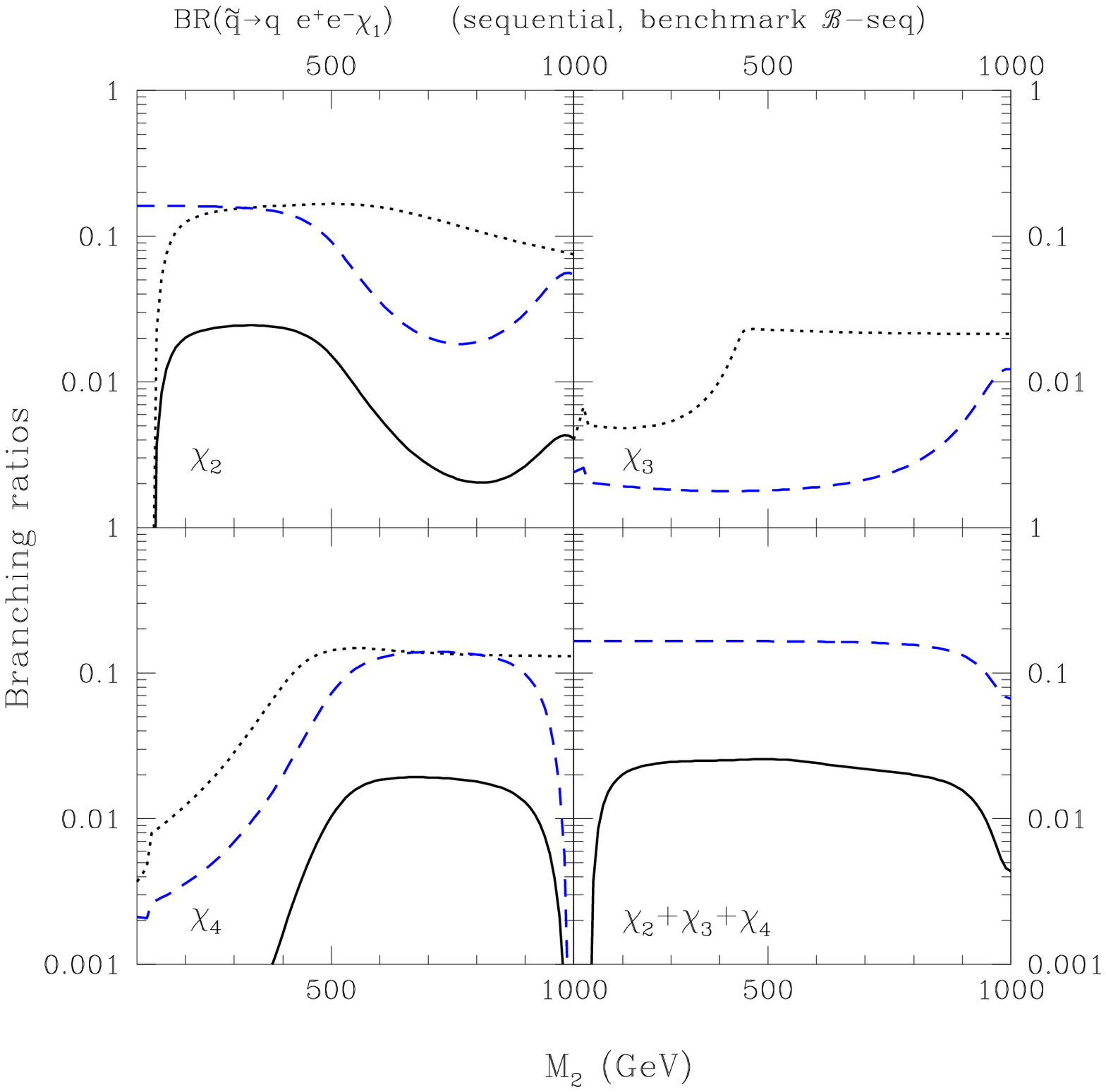}
\end{center}
\caption{Branching ratios for the sequential process  
$\tilde{q} \rightarrow \bar{e} e  \chi_1$
 in benchmark $\mathcal{B}$--seq
as functions of $M_2$. Notations are as in Fig.\protect\ref{fig:aseq2_e}.
\label{fig:bseq_e}}
\end{figure}

\begin{figure}
\begin{center}
\includegraphics[width=12cm]{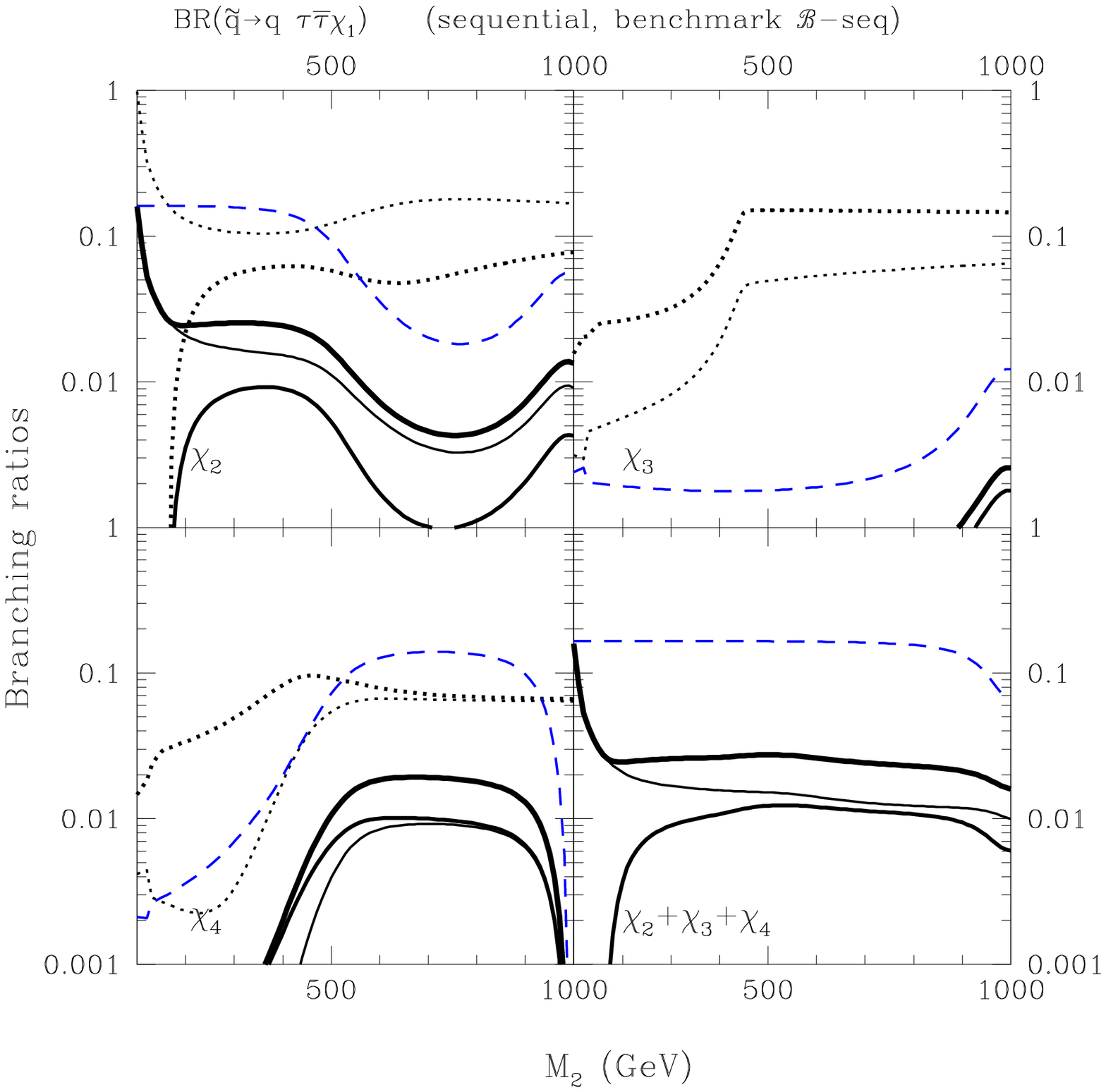}
\end{center}
\caption{Branching ratios for the sequential process  
$\tilde{q} \rightarrow \bar{\tau} \tau  \chi_1$
 in benchmark $\mathcal{B}$--seq
as functions of $M_2$. Notations are as in Fig.\protect\ref{fig:aseq1_tau}.
\label{fig:bseq_tau}}
\end{figure}

The scenario $\mathcal{B}$ is characterized by a heavy pseudoscalar Higgs mass
($m_A\gsim$ 200 GeV) and by a very light stau, which is required in
order to keep the neutralino relic abundance below the observational
limit.  Moreover, as explained in Section
\ref{sec:cosmological_properties}, $|\mu|$ must be large. We define in
this case the following benchmark, corresponding to the lightest
possible mass of the neutralino $\chi_1$:

\begin{eqnarray}
{\bf \mathcal{B}{\rm -seq}} &:& M_1=25\; {\rm GeV}\;\;\;\; \mu=-500 \;{\rm
  GeV}\;\;\;\;\tan\beta=10\nonumber\\
 &  &   m_{\tilde{\tau}}=87\;{\rm GeV}
\;\;(m_{\tilde{l}}=120\;{\rm GeV})\;\;A=0.
\end{eqnarray}

\noindent
This benchmark is depicted qualitatively in Fig.\ref{fig:sequential}
and summarized in Table \ref{table:benchmarks}.  The branching ratios
for this benchmark are shown in Figs. \ref{fig:bseq_e}--\ref{fig:bseq_tau}. 
In this case the
possible hierarchy between $M_2$ and $|\mu|$ is richer than in the
previous benchmark $\mathcal{A}$-seq1, since now also $M_2 << |\mu|$
can occur. In particular, this implies that the compositions of
$\chi_2$ and $\chi_4$ flip the one into the other in going from
$M_2<|\mu|$ to $M_2>|\mu|$ (an example of this behaviour can be seen
in Fig.\ref{fig:neutralino_compositions}, although for a slightly
different set of supersymmetric parameters).  Here, for $M_2 <$ 500
GeV, $\chi_2$ is mainly a gaugino, whereas $\chi_4$ is dominantly a
Higgsino; the other way around, for $M_2 >$ 500 GeV. While $\chi_2$
and $\chi_4$ exchange their roles as $M_2$ runs over its range,
$\chi_3$ is steadily a Higgsino, independently of $M_2$. On the basis
of these properties one understands the features of the branching
ratios for the production of the $\chi_i$ intermediate states (dashed
lines).  It is also clear why their combined results (dashed line in
the bottom-right panel) have a much milder dependence on $M_2$.

As for the peculiar behaviour of the branching ratios for the decays
$\chi_{i}\rightarrow l\bar{l}\chi_1$ (dotted lines), notice that
their sudden drop at low $M_2$ in the cases of $\chi_3$ and $\chi_4$
is due to the opening of some competing decay channels. In fact, for $M_2 <
|\mu|$ one has $m_3 \simeq m_4 \simeq |\mu|$ and, at the same time,
the chargino mass is of order $M_2$. Thus, under these circumstances,
$\chi_3$ and $\chi_4$ have a sizable branching ratio into a
chargino-$W$ state ($\sim$ 54\% at $M_2$ = 300 GeV). In addition, also
the channel $\chi_3 \rightarrow \chi_2 Z$ becomes important in this
case (with a branching ratio of about $22 \%$ at $M_2$ = 300 GeV).

\subsection{Branched chain benchmarks}
\label{sec:branched}

\begin{figure}[t]
\includegraphics[width=5.5cm,bb= 0 340 590 540,clip=true]{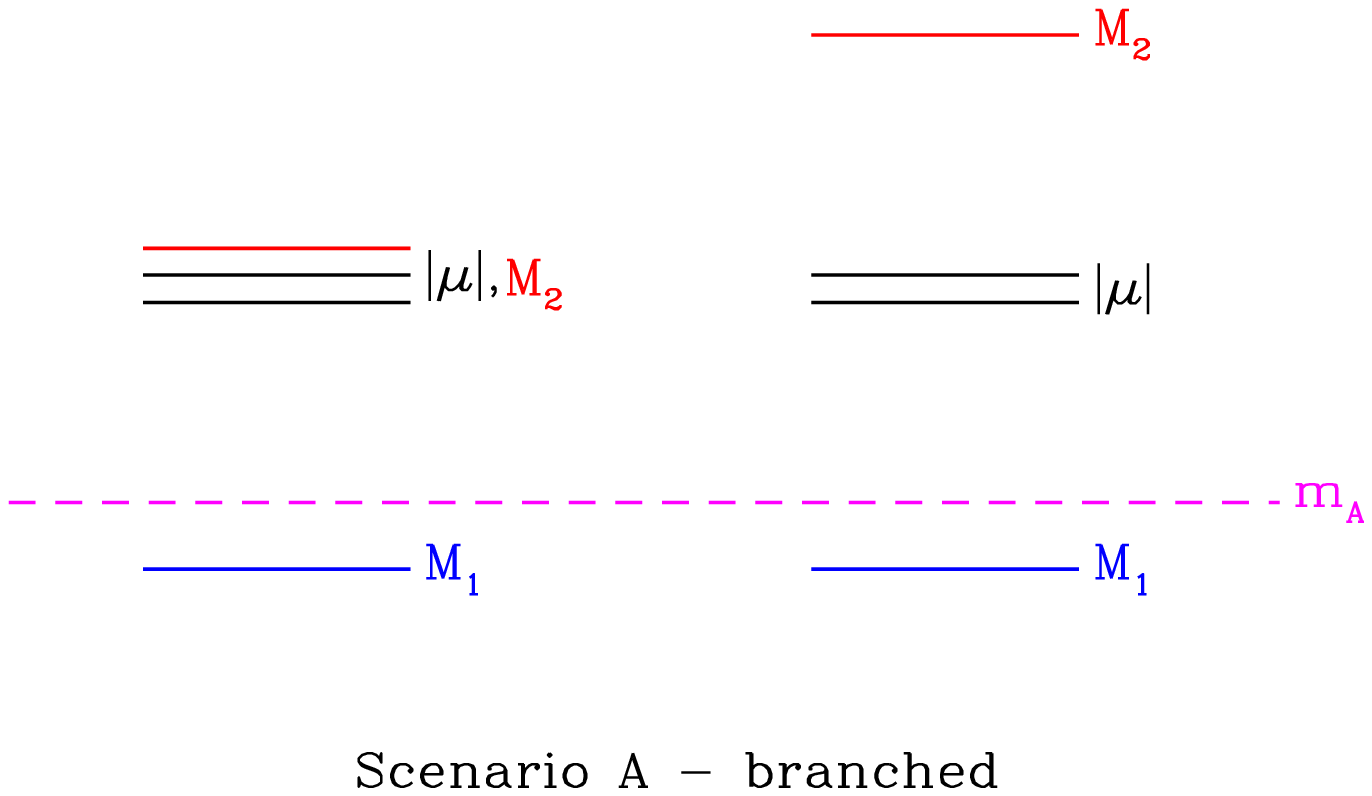}
\includegraphics[width=5.5cm,bb= 0 340 590 540,clip=true]{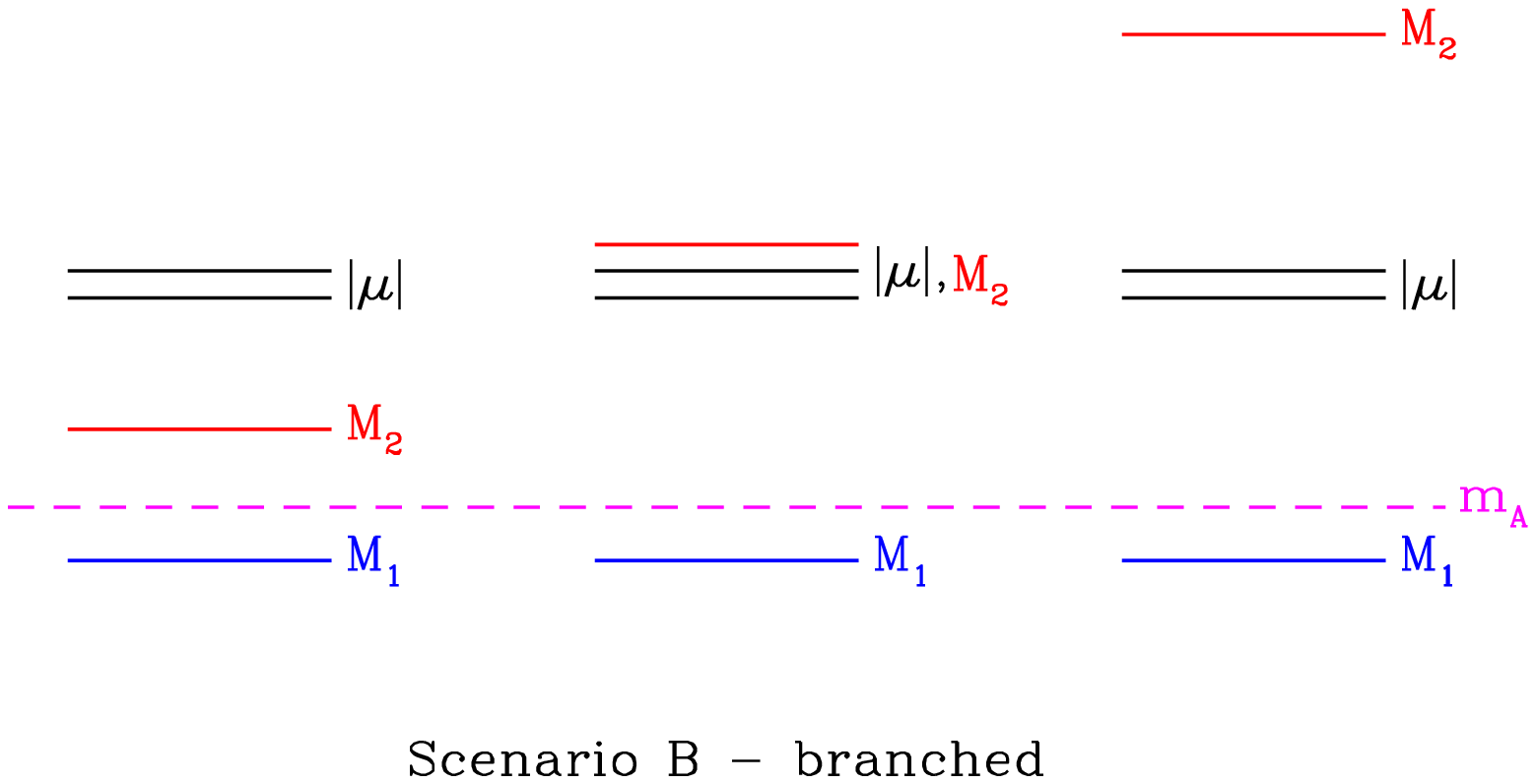}
\includegraphics[width=5.5cm,bb= 0 340 590 540,clip=true]{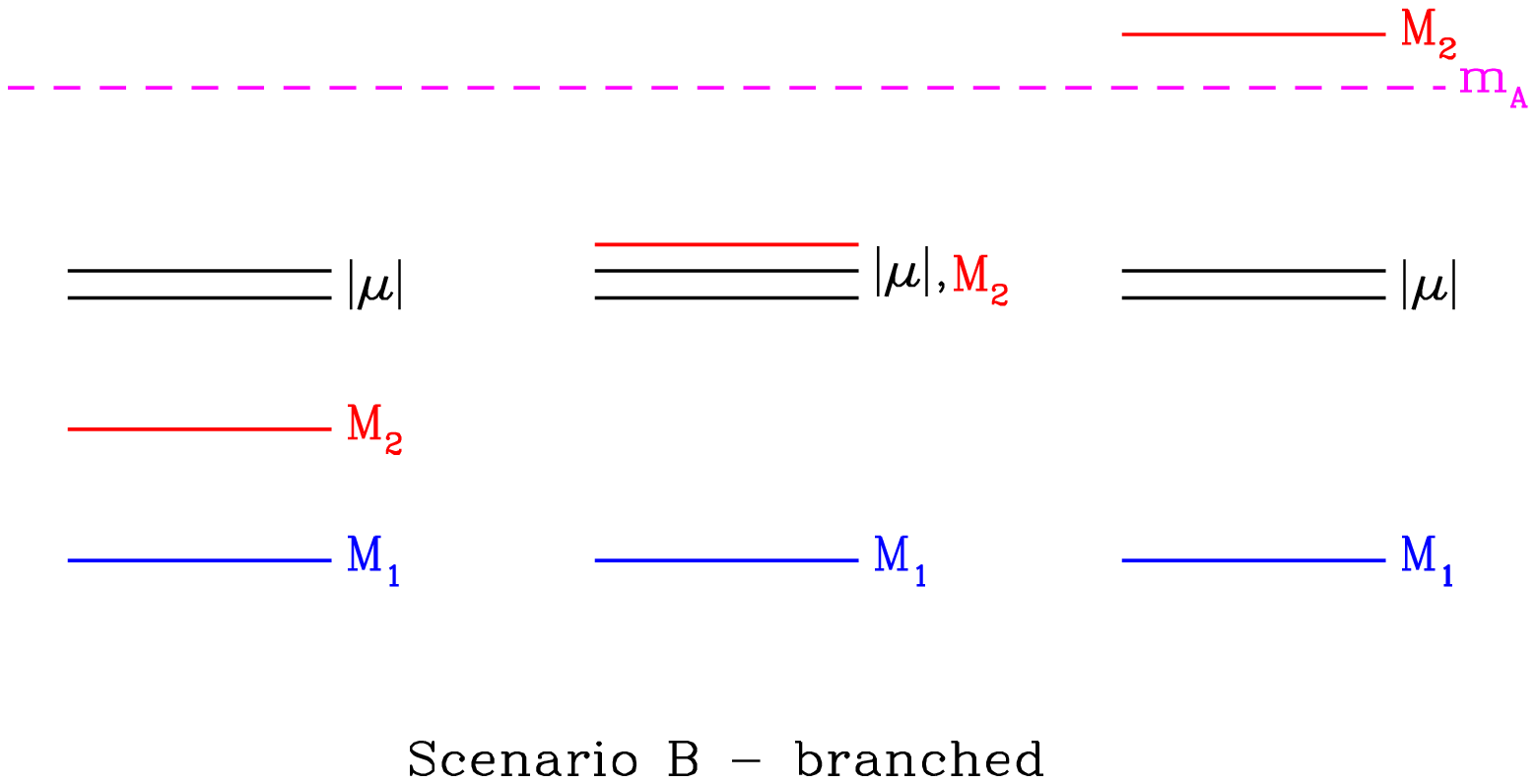}
\caption{Qualitative schemes of benchmarks for branched decay
chains: $\mathcal{A}$--brc, $\mathcal{B}$--brc1, $\mathcal{B}$--brc2. For each benchmark, extremes values for $M_2$
are considered: $M_2 \sim |\mu|$ and $M_2 > |\mu|$ for scenario $\mathcal{A}$;
$M_2 < |\mu|$, $M_2 \sim |\mu|$ and $M_2 > |\mu|$ for scenario $\mathcal{B}$.
\label{fig:branched}.}
\end{figure}

\begin{figure}
\begin{center}
\includegraphics[width=12cm]{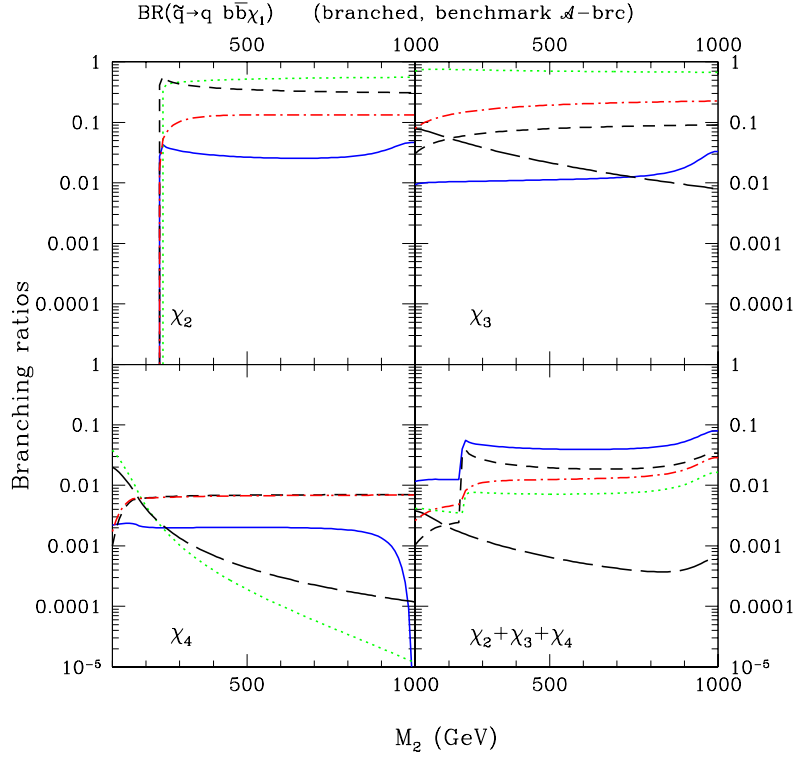}
\end{center}
\caption{Branching ratios for the branched chain in benchmark $\mathcal{A}$--brc
as functions of $M_2$. Each of the first three panels corresponds to
a different intermediate neutralino state ($\chi_i, i=2,3,4$) in the
branched chain, while the last panel provides the  branching
ratios summed over all $\chi_i$'s. Notations are as follows.  Dotted
lines: branching ratio for the process $\chi_{i}\rightarrow Z\chi_1$;
short--dashed lines: branching ratio for the process
$\chi_{i}\rightarrow h\chi_1$; long--dashed lines: branching ratio for
the process $\chi_{i}\rightarrow H\chi_1$; dot-dashed lines: branching
ratio for the process $\chi_{i}\rightarrow A\chi_1$; solid line:
combined branching ratios for the whole decay chain
$\tilde{q}\rightarrow q \chi_{i}\rightarrow q (Z,h,H,A)
\chi_1\rightarrow q b\bar{b}\chi_1$. The branching ratios for the
process $\tilde{q}\rightarrow q \chi_i$, already displayed in Figs.
\protect\ref{fig:aseq2_e},\protect\ref{fig:aseq2_tau}, are omitted here.
\label{fig:abrc}}
\end{figure}

\begin{figure}
\begin{center}
\includegraphics[width=12cm]{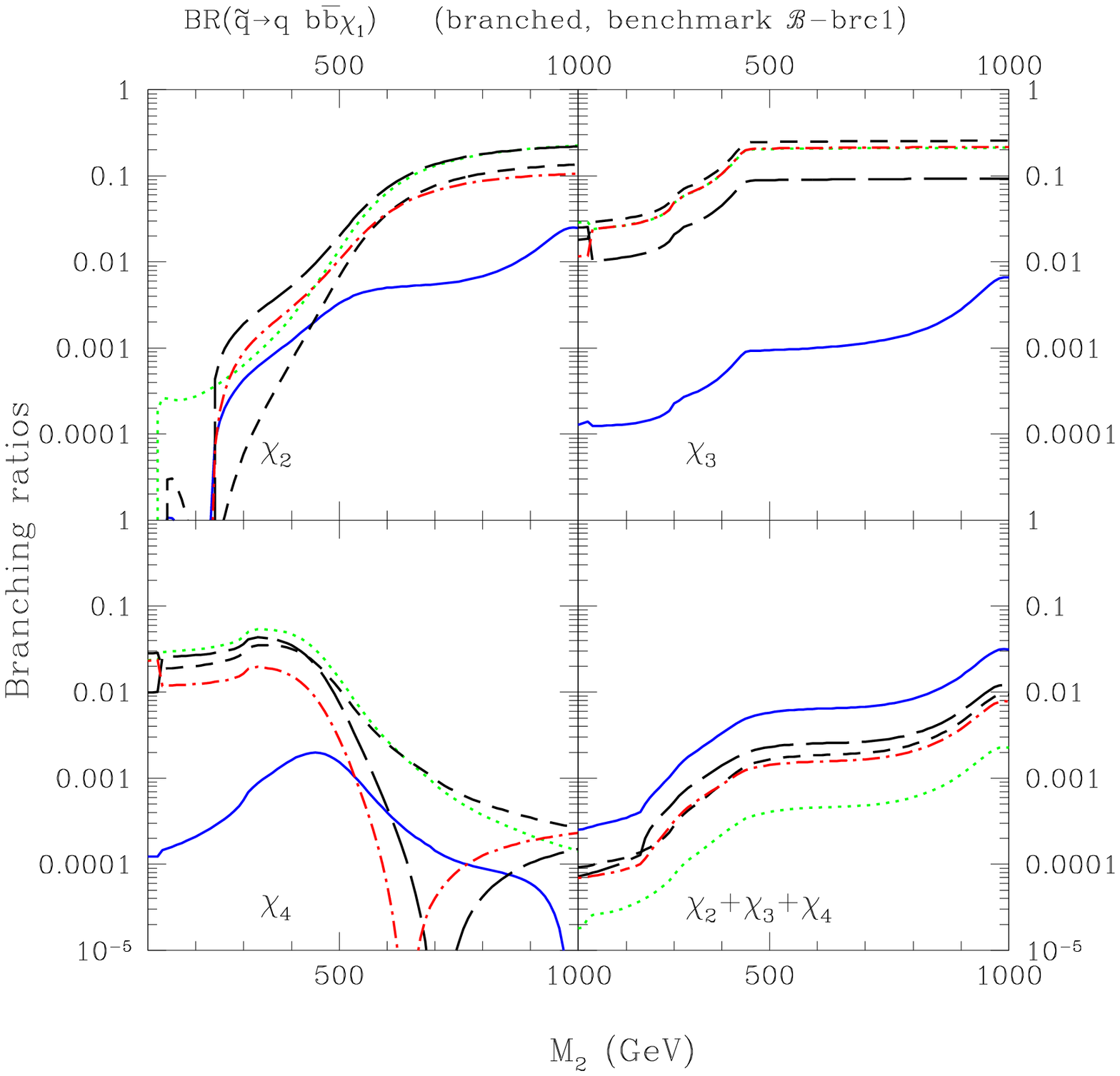}
\end{center}
\caption{
Branching ratios for the branched chain in benchmark $\mathcal{B}$--brc1
as functions of $M_2$.
Notations are as in Fig.\protect\ref{fig:abrc}.
The branching ratios for the
process $\tilde{q}\rightarrow q \chi_i$, already displayed in Figs.
\protect\ref{fig:bseq_e},\protect\ref{fig:bseq_tau}, are omitted here.
\label{fig:bbrc1}}
\end{figure}

In the branched chains the decay amplitude is sensitive to the value of $m_A$, while it
does not depend on $m_{\tilde{l}}$. As a consequence of this, as far
as scenario $\mathcal{A}$ is considered, for branched decays we adopt the
following benchmark:

\begin{eqnarray}
{\bf \mathcal{A}{\rm -brc}}&: & M_1=10\; {\rm GeV}\;\;\;\; \mu=110 \;{\rm
  GeV}\;\;\;\;\nonumber\\
       &          & \tan\beta=35\;\;\;\;m_{A}=90\;{\rm GeV}\;\;\;\;A=0,
 \label{eq:A-brc}
\end{eqnarray}


\noindent while the value of $m_{\tilde{l}}$ is undetermined (for
definiteness, in this case we fix $m_{\tilde{l}}=150$ GeV).  This
benchmark is depicted qualitatively in Fig.\ref{fig:branched} and
summarized in Table \ref{table:benchmarks}.  The branching ratios for
benchmark $\mathcal{A}$--brc are displayed in Fig.\ref{fig:abrc},
where the dotted lines show the branching ratio for the process
$\chi_{i}\rightarrow Z\chi_1$, the short--dashed lines correspond to
the process $\chi_{i}\rightarrow h\chi_1$, the long--dashed lines
indicate the process $\chi_{i}\rightarrow H\chi_1$, the dot-dashed
lines denote the process $\chi_{i}\rightarrow A\chi_1$ and, finally,
the solid line indicates the combined branching ratio for the whole
decay chain $\tilde{q}\rightarrow q \chi_{i}\rightarrow q (Z,h,H,A)
\chi_1\rightarrow q b\bar{b}\chi_1$. Note that in this Figure the
branching ratios for the process $\tilde{q}\rightarrow q \chi_i$,
already displayed in Figs. \ref{fig:aseq1_e},\ref{fig:aseq1_tau} for
the same set of parameters, are omitted. The combined branching ratio
for the decay $\tilde{q}\rightarrow q \chi_{i}\rightarrow q Z
\chi_1\rightarrow q e^+e^-\chi_1$, to which only the $Z$ exchange
contributes, can be simply calculated by scaling the line for $Z$ by a
factor 0.22, the ratio of $BR(Z\rightarrow e^+e^-)$ and
$BR(Z\rightarrow b\bar{b})$ in the Standard Model.

As shown in Fig.\ref{fig:abrc}, the major contribution to the
branching ratio BR($\tilde{q} \rightarrow b \bar{b} \chi_1$) is due to
the production of $\chi_2$, which subsequently decays into $\chi_1$
through the production of a $Z$ boson or a $A$ Higgs boson, provided
$|m_2|$ is above threshold (which implies that $M_2 \gsim$ 150 GeV).
These two channels have large branching ratios, due to the fact that
$\chi_2$ is mainly a Higgsino, and also $\chi_1$ has a sizable
Higgsino component.  The branching ratio of the channel through $Z$
prevails over the branching ratio of the channel through $A$ roughly
by a factor 4, since a factor 7-10 due to the Lorentz structure of the
matrix elements is partially compensated by a factor 0.7-0.5 due to the
different coupling constants.

\begin{figure}
\begin{center}
\includegraphics[width=12cm]{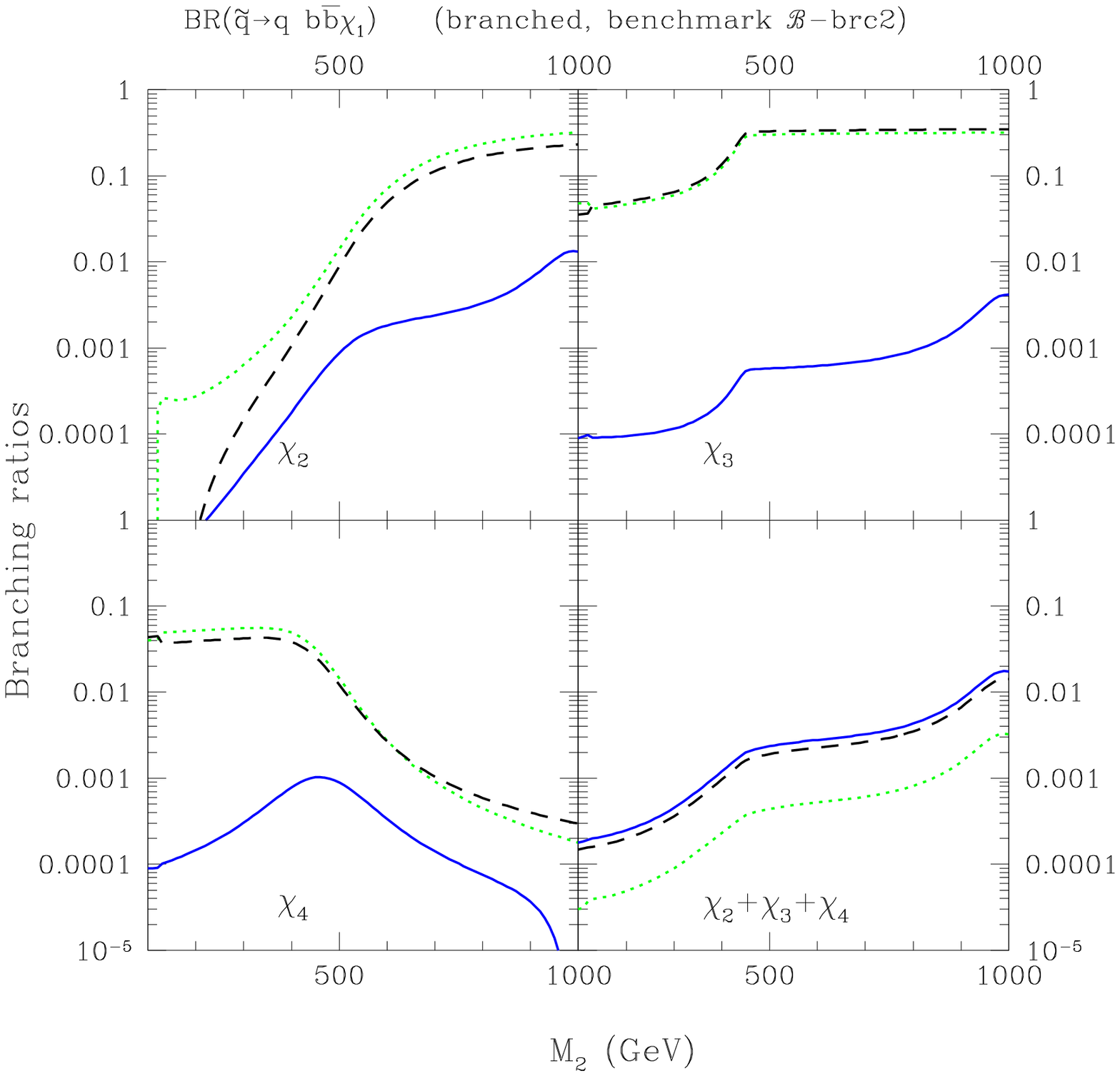}
\end{center}
\caption{
Branching ratios for the branched chain in benchmark $\mathcal{B}$--brc2
as functions of $M_2$.
Notations are as in Fig.\protect\ref{fig:abrc}.
The branching ratios for the
process $\tilde{q}\rightarrow q \chi_i$, already displayed in Figs.
\protect\ref{fig:bseq_e},\protect\ref{fig:bseq_tau}, are omitted here.
\label{fig:bbrc2}}
\end{figure}

As far as branched decays are concerned, the most relevant feature of
the scenario $\mathcal{B}$ is that  $m_A$ is heavy ($m_A\gsim$ 200 GeV),  while
$\tan\beta\lsim 10$ is moderate. As a consequence of this, lower
branching ratios are expected compared to the previous case of
benchmark $\mathcal{A}$--brc. In scenario $\mathcal{B}$ we adopt the two following
benchmarks:

\begin{eqnarray}
{\bf \mathcal{B}{\rm -brc1:}} & m_{A}={\rm 200 \; GeV}
\label{eq:B-brc1}\\
{\bf \mathcal{B}{\rm -brc2:}} & m_{A}=1000 \;{\rm GeV},
\label{eq:B-brc2}
\end{eqnarray}
\noindent where, in both cases, $\mu=-500$ GeV, $M_1$=25 GeV,
$\tan\beta$=10, $m_{\tilde{l}}=120$  GeV and $A=0$.
These benchmarks are depicted qualitatively in Fig.\ref{fig:branched}
and summarized in Table \ref{table:benchmarks}.
The corresponding branching ratios are shown in Figs.\ref{fig:bbrc1} and
\ref{fig:bbrc2}. Note that in Fig.\ref{fig:bbrc2} only the decays
$\chi_{i}\rightarrow Z\chi_1$ and $\chi_{i}\rightarrow h\chi_1$ are
kinematically allowed.  The main features of these figures may readily be
accounted for by arguments similar to the ones previously described.

\begin{figure}
\begin{center}
\includegraphics[width=12cm]{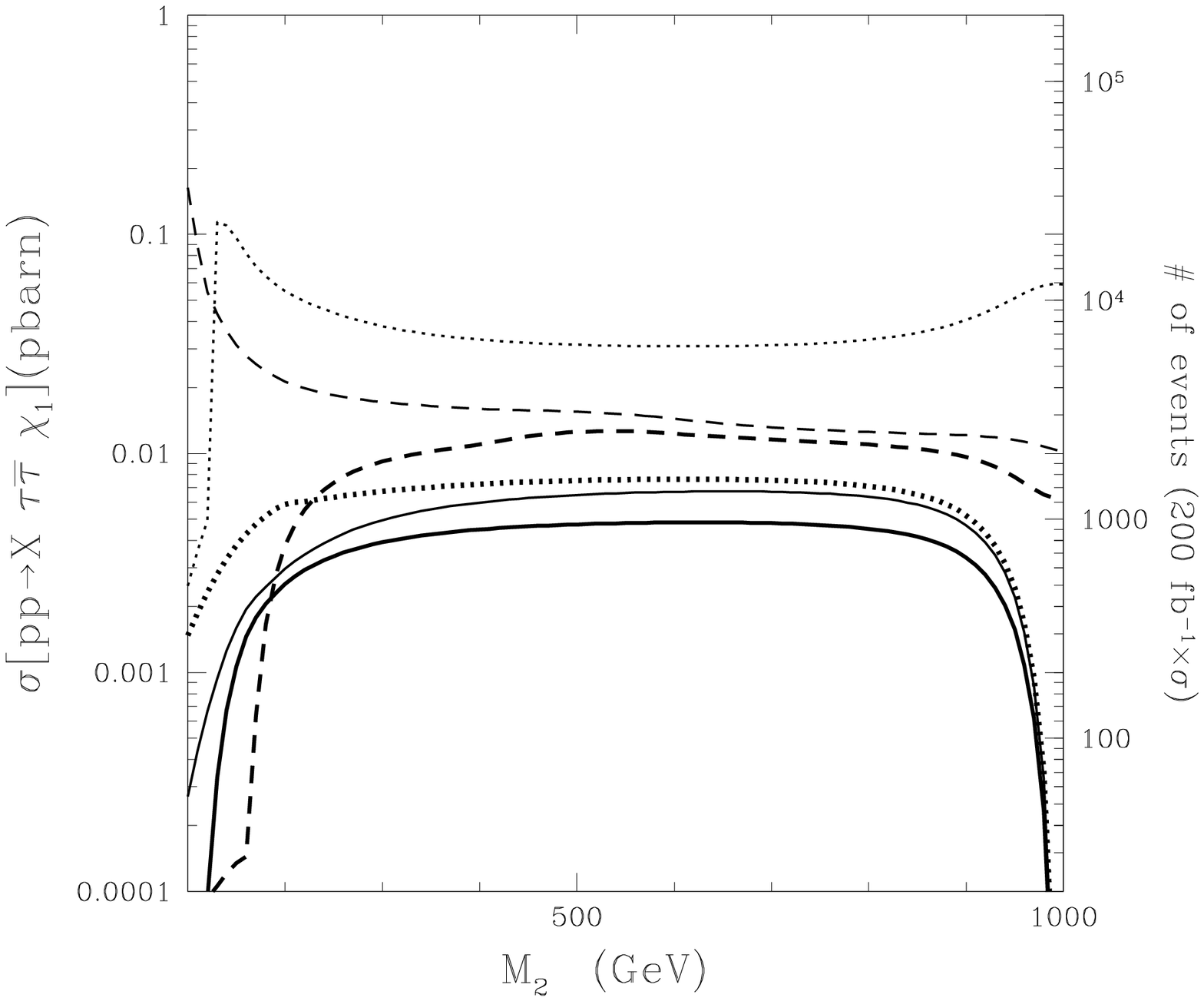}
\end{center}
\caption{Total production cross sections for the processes
$pp\rightarrow X \tau \bar{\tau}\chi_1$ in the benchmarks for sequential chains
for $\sqrt{s}=14$ TeV, as functions of $M_2$.
The decay branching ratios included in the
calculation are those displayed in
Figs. \protect\ref{fig:aseq1_tau},\protect\ref{fig:aseq2_tau},\protect\ref{fig:bseq_tau}.  The notation is as
follows. 
Thin--solid line: $\mathcal{A}$--seq1 with $m_{\tilde{l}}=150$ GeV 
and mediation of $\tilde{\tau}_1$;  
thick--solid line: $\mathcal{A}$--seq1 with $m_{\tilde{l}}=150$ GeV 
and mediation of $\tilde{\tau}_2$; 
thin--dotted line: $\mathcal{A}$--seq2 with mediation of $\tilde{\tau}_1$; 
thick--dotted line: $\mathcal{A}$--seq2 with mediation of $\tilde{\tau}_2$;
thin--dashed line: $\mathcal{B}$--seq with mediation of $\tilde{\tau}_1$;
thick--dashed line: $\mathcal{B}$--seq with mediation of $\tilde{\tau}_2$.
On the right
vertical axis the corresponding total number of expected events is shown, assuming a
luminosity of 200 fb$^{-1}$ (two years of high--luminosity run at
LHC).
\label{fig:cross_sections_sequential_tau}}
\end{figure}

\begin{figure}
\begin{center}
\includegraphics[width=12cm]{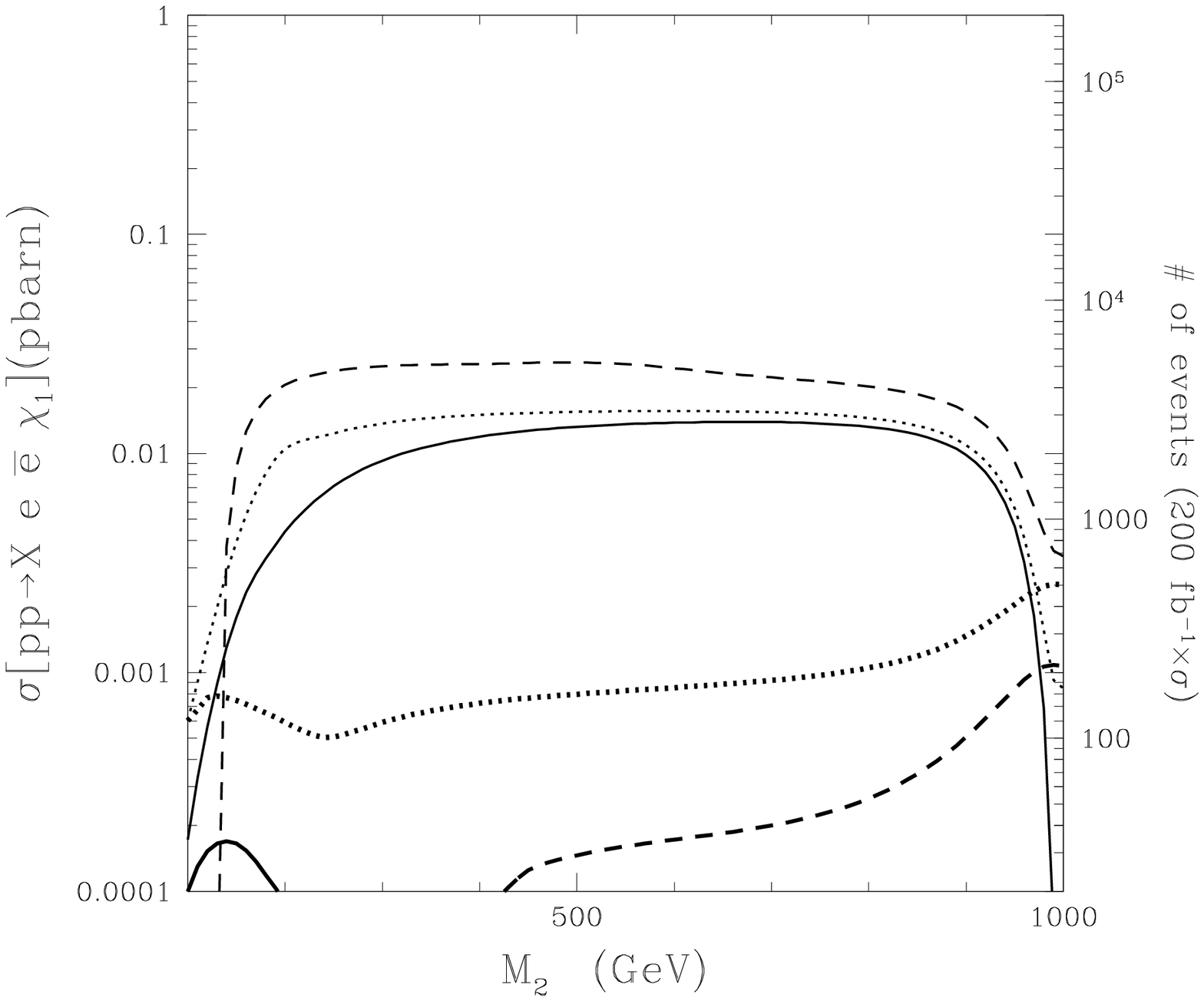}
\end{center}
\caption{Total production cross sections for the processes
$pp\rightarrow X e \bar{e}\chi_1$ in the benchmarks for sequential
chains for $\sqrt{s}=14$ TeV, as functions of $M_2$.  The decay branching
ratios included in the calculation are those displayed in
Figs. \protect\ref{fig:aseq1_e},\protect\ref{fig:aseq2_e},\protect\ref{fig:bseq_e}.
The notation is as follows.  Thin--solid line: $\mathcal{A}$--seq1
with $m_{\tilde{l}}=150$ GeV and mediation of $\tilde{e}_L$;
thick--solid line: $\mathcal{A}$--seq1 with $m_{\tilde{l}}=150$ GeV
and mediation of $\tilde{e}_R$; thin--dotted line: $\mathcal{A}$--seq2
with mediation of $\tilde{e}_L$; thick--dotted line:
$\mathcal{A}$--seq2 with mediation of $\tilde{e}_R$; thin--dashed
line: $\mathcal{B}$--seq with mediation of $\tilde{e}_L$;
thick--dashed line: $\mathcal{B}$--seq with mediation of
$\tilde{e}_R$.  On the right vertical axis the corresponding total
number of expected events is shown, assuming a luminosity of 200
fb$^{-1}$ (two years of high--luminosity run at LHC).
\label{fig:cross_sections_sequential_e}}
\end{figure}

\begin{figure}
\begin{center}
\includegraphics[width=12cm]{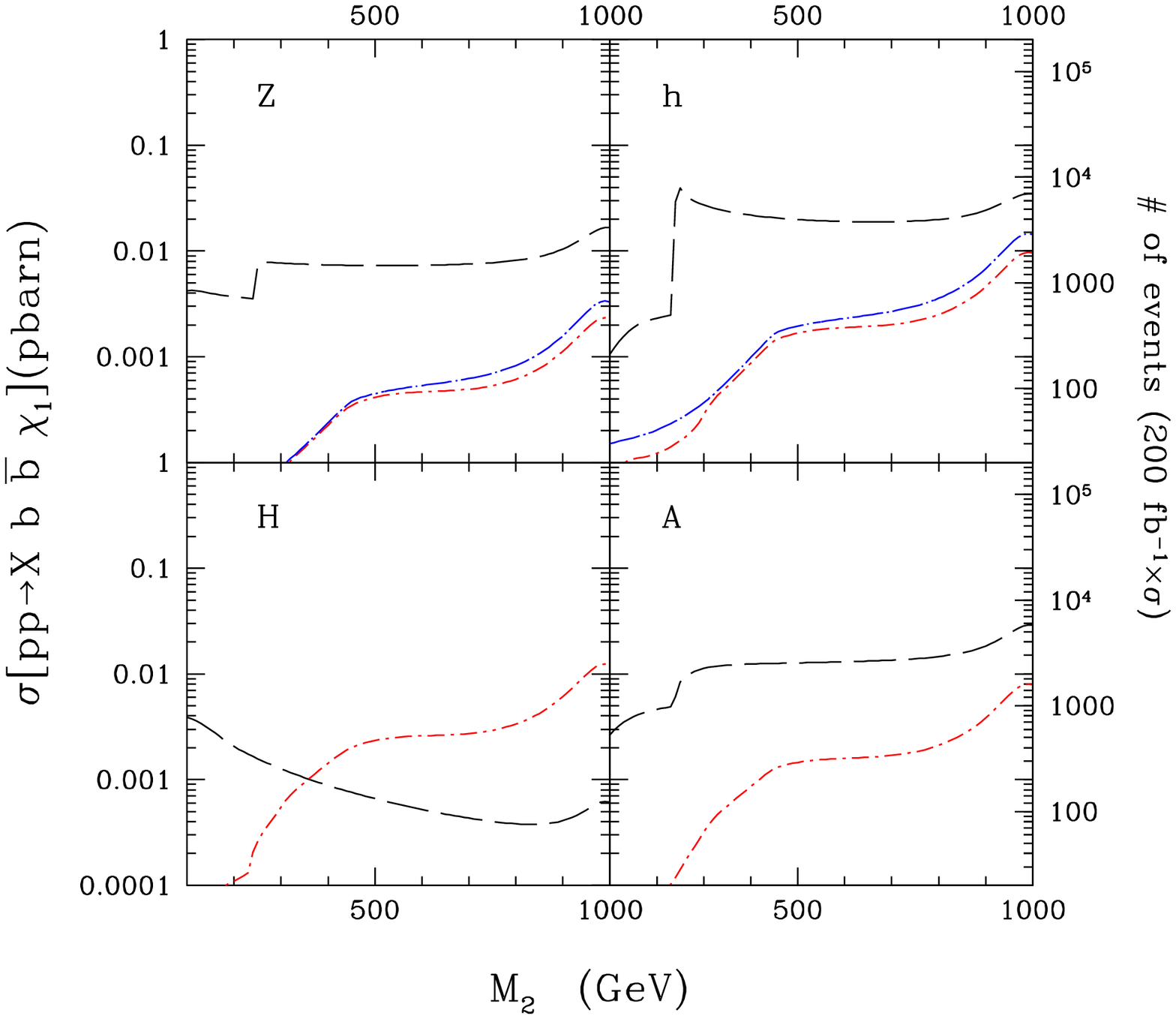}
\end{center}
\caption{Total production cross sections for the processes
$pp\rightarrow X b \bar{b}\chi_1$ in the benchmarks for branched chains,
discussed in the
text, and for $\sqrt{s}=14$ TeV, as functions of $M_2$.
The decay branching ratios included in the
calculation are those displayed in
Figs. \protect\ref{fig:abrc}--\protect\ref{fig:bbrc2}.
The top-left panel refers to the production of a $Z$-boson,
the top-right panel to the production of an $h$-boson,
the bottom-left panel  to the production of an $H$-boson, and
the bottom-right panel to the production of an $A$-boson.
 The notation for the curves is as
follows. Long--dashed line: $\mathcal{A}$--brc;
dot--short dashed line: $\mathcal{B}$--brc1; dot--long dashed line:
$\mathcal{B}$--brc2. On the right vertical axis
the corresponding total number of expected events is shown, assuming a
luminosity of 200 fb$^{-1}$ (two years of high--luminosity run at
LHC).
\label{fig:cross_sections_branched}}
\end{figure}

\subsection{Expected number of events}
\label{sec:expected}

In order to assess the experimental detectability of our scenarios, we
now display in
Figs.\ref{fig:cross_sections_sequential_tau} --
\ref{fig:cross_sections_branched} the total production cross sections for
the process $pp\rightarrow X f\bar{f}
\chi_1$ with $\sqrt{s}=14$ TeV for each of the benchmarks previously
introduced ($f=\tau$ and $f=e$, for the cases $\mathcal{A}$--seq1, 
$\mathcal{A}$--seq2 and $\mathcal{B}$--seq, are 
displayed in Fig. \ref{fig:cross_sections_sequential_tau} and in 
Fig. \ref{fig:cross_sections_sequential_e}, respectively; 
 $f=b$, for the cases $\mathcal{A}$--brc, $\mathcal{B}$--brc1 and 
 $\mathcal{B}$--brc2, is 
displayed in Fig. \ref{fig:cross_sections_branched}). 
 Notice that the case $f = e$ for branched chains can simply be derived by 
 scaling  the lines of the top-left panel of 
 Fig. \ref{fig:cross_sections_branched}
 (production of a  $Z$-boson) 
 by a factor 0.22,  the ratio of $BR(Z\rightarrow e^+e^-)$ and 
      $BR(Z\rightarrow b\bar{b})$ in the Standard Model.

  A rough estimate of the expected
events after two year's high--luminosity running at LHC is indicated
on the vertical axis on the right of these figures.
For the calculation of the production cross sections
$\sigma(pp\rightarrow
\tilde{q}\tilde{q},\tilde{q}\tilde{q}^*,\tilde{g}\tilde{g},
\tilde{q}\tilde{g})$ we have used the code PROSPINO \cite{prospino}

We assume here arbitrarily that the cuts required to extract
the SUSY signal from the background will have an efficiency 
of the order of 10\%, and a reasonably precise measurement
can be performed for a chain for which $\sim 100$ events are left 
after the experimental cuts.

From the results summarized in
Figs.\ref{fig:cross_sections_sequential_tau} --
\ref{fig:cross_sections_branched} it turns out that there are very
good perspectives for a fruitful investigation of the supersymmetric
parameter space relevant for light neutralinos at LHC. In particular
we notice that: 1) The scenario $\mathcal{A}$ should be easily
explorable both through sequential decay chains and through branched
chains.  2) For the sequential case good perspectives are offered by
the $\bar{e} e$ signal and by the $\bar{\tau} \tau$ one in both
benchmarks $\mathcal{A}$--seq1 and $\mathcal{A}$--seq2. In particular,
notice that in the benchmark $\mathcal{A}$--seq1 the $\bar{e} e$ and
the $\bar{\tau} \tau$ signals are about of the same size, because they
take origin from the decay of $\chi_4$, which for $M_2 > |\mu|$ is
dominantly a gaugino; on the contrary, in the benchmark
$\mathcal{A}$--seq2 the $\bar{\tau} \tau$ signal is larger than the
$\bar{e} e$ one, since in this case the process goes mainly through
intermediate Higgsino-like neutralino states.  3) For the branched
case the scenario $\mathcal{A}$ gives good measurement perspectives
for production of a $\bar{b} b$ pair through Z and h, A Higgs bosons.
4) In scenario $\mathcal{B}$ large signals are expected in terms of
$\bar{e} e$ and $\bar{\tau} \tau$ pairs in sequential processes; these
two signals are comparable in size, since they are generated by
intermediate neutralino states which are dominantly gauginos.  5)
Higher statistics will be required to explore the scenario
$\mathcal{B}$ by branched decay; the most favorable processes are
represented by those mediated by Z or h, A Higgs bosons when $M_2 >
|\mu|$.

The rates are of course function of the assumed squark mass of 
1~TeV. Since the BR for the considered chains do not depend
on the squark mass, results for different squark masses 
can be obtained by scaling down the curves by the relative 
squark production cross-section. As the squark mass gets nearer 
to the gluino mass a significant contribution will also come
from squarks produced in gluino decays.
For instance for a gluino mass of 2 TeV and a squark mass
of $\sim 1500 (1900) $~GeV, the scaling factor would be 
respectively $\sim 6 (25)$, and the fraction of events with at least 
a gluino in the initial state would be respectively $\sim 18\% (55\%)$.

We stress that here our considerations are simply based on the
evaluated total number of events. The actual potentiality of
investigation at LHC of the present signals will require a detailed
analysis in terms of signal to background ratios and specific
kinematical distributions. This further investigation is beyond the
scope of this paper and will be presented elsewhere \cite{future}.

\section{Conclusions}
\label{sec:conclusions}

In this paper we have analyzed the discovery potential of LHC with
respect to light neutralinos, {\it i.e.} neutralinos with a mass
$m_{\chi} \lsim 50$ GeV, which arise in supersymmetric models where
gaugino masses are not unified at a Grand Unified (GUT) scale.  These
neutralinos have been thoroughly investigated in
Refs. \cite{lowneu,lowdir}, under the hypotheses that R-parity is
conserved and that the lowest neutralino state is the Lightest
Supersymmetric Particle (LSP). This LSP light neutralino has been
proved to be quite interesting as a cold dark matter particle
intervening in a number of direct and indirect astrophysical effects.

In particular, in  Refs. \cite{lowneu,lowdir} it was derived that 
the present constraints due to accelerator and other precision measurements,
together with limits imposed by cosmological observations, concur in confining
these  neutralinos to configurations located in 
a well delimited part of the  supersymmetric parameter space. 
In other words, the relevant region of the SUSY parameter space can be described
by a limited number of free parameters, since some of the model parameters are
essentially frozen by the external constraints. 

Such a situation implies that a few scenarios and relevant conspicuous
benchmarks can  naturally  be singled out. 
This is the strategy that we have used in the present paper in order to explore 
the discovery potential of LHC as far as light neutralinos are concerned. 

The simplicity of the underlying supersymmetric model also allows the derivation of
analytic formulae which help a lot in understanding the main properties of the
spectroscopy of the four neutralino states. The relevant expressions have been
derived in the first part of the paper. 

Two main scenarios have been introduced: the scenario  $\mathcal{A}$, where the
stable neutralino has a mass $m_{\chi} \sim$ 10 GeV, and the scenario 
$\mathcal{B}$ where $m_{\chi} \sim$ 25 GeV (the specification for the values 
of the other supersymmetric
parameters is given in Table \ref{table:scenarios}). Within these two scenarios a
number of convenient benchmarks have been introduced 
(see Table \ref{table:benchmarks}). 

In the framework of the selected benchmarks we have considered the 
following decay chains, generated by squarks produced in the proton--proton
scattering: 
$\tilde{q}\rightarrow q \chi_i\rightarrow q \tilde{f}f\rightarrow q
\bar{f}f \chi_1$ (sequential chain), and 
$\tilde{q}\rightarrow q \chi_i\rightarrow q (Z,h,H,A) \chi_1\rightarrow q
\bar{f}f \chi_1$ (branched chain). 
We limited our discussion to the case in which the gluino
is heavier than the squark; for definiteness, we have set
the SU(3) gaugino mass at the representative value $M_3$ = 2 TeV and
the squark soft-mass at the
 value $m_{\tilde{q}}$ = 1 TeV.  Notice that these two parameters
are irrelevant in the specification of our scenarios inspired by cosmology.

Branching ratios and number of events expected at LHC have been evaluated for
the  signals which proved to be the most important ones
for experimental investigation. 

We arrived at the following main conclusions: 

i) The scenario $\mathcal{A}$ should be easily explorable both through 
 sequential decay chains and through branched chains.
  For the sequential case good perspectives are offered by the 
 $\bar{e} e$ (or $\bar{\mu} \mu$) signal and by the $\bar{\tau} \tau$ one. 
 For the branched case the scenario  $\mathcal{A}$ gives good measurement 
perspectives for production of a $\bar{b} b$ pair through Z and h, A Higgs
bosons, combined with light lepton pairs through Z.  

ii) In scenario $\mathcal{B}$ large signals are expected in terms of 
$\bar{e} e$  (or $\bar{\mu} \mu$) 
 and  $\bar{\tau} \tau$ pairs in sequential processes. 
High statistics will be required to explore the scenario  $\mathcal{B}$ 
by branched decay; the most favorable processes being represented by those 
mediated by Z or h, A Higgs bosons when $M_2 > |\mu|$. 

These results show that LHC has a strong potential in the
investigation  of the supersymmetric parameter region compatible 
with a light neutralino. Due to the characteristic features of this region, the
measurements of LHC should easily prove or disprove our model. 

Finally, we wish to recall that our conclusions  are based on the 
evaluated total number of events, only. To ascertain the actual potentiality 
of LHC in the study of our model, the investigation has to be pursued 
to include  analysis of the signal/background ratios and of specific 
kinematical distributions. This further investigation will be presented in a
subsequent publication \cite{future}.

\acknowledgements We acknowledge Research Grants funded jointly by
Ministero dell'Istruzione, dell'Universit\`a e della Ricerca
(MIUR), by Universit\`a di Torino and by Istituto Nazionale di
Fisica Nucleare within the {\sl Astroparticle Physics Project}.


\begin{thebibliography}{99}


\bibitem{lowneu}
A. Bottino, N. Fornengo and S. Scopel, Phys. Rev. D
{\bf 67}, 063519 (2003)
;
A. Bottino, F. Donato, N. Fornengo and S. Scopel,
 Phys. Rev. D {\bf 68}, 043506 (2003).

\bibitem{lowdir} A. Bottino, F. Donato, N. Fornengo and S. Scopel: 
 Phys. Rev. D {\bf 69}, 037302 (2004); 
 Phys. Rev. D {\bf 70}, 015005 (2004) and 
 Phys. Rev. D {\bf 77}, 015002 (2008).

\bibitem{atltdr} ATLAS Collaboration, {\it ATLAS detector and
physics performance Technical Design Report},
CERN/LHCC 99-14/15 (1999).
http://atlas.web.cern.ch/Atlas/GROUPS/PHYSICS/TDR/access.html.

\bibitem{cmstdr2}
CMS Collaboration {\em CMS physics : Technical Design Report v.2 : Physics performance}
CERN-LHCC-2006-021 (2006).
http://cdsweb.cern.ch/search.py?recid=942733.




\bibitem{LEPb} A. Colaleo (ALEPH Collaboration), talk at
SUSY'01, June 11-17, 2001, Dubna, Russia; J. Abdallah et al.
(DELPHI Collaboration), DELPHI 2001-085 CONF 513, June 2001;
LEP Higgs Working Group for Higgs boson searches, arXiv:hep-ex/0107029;
LEP2 Joint SUSY Working Group, {\tt http://lepsusy.web.cern.ch/lepsusy/}.

\bibitem{cdf} A.A. Affolder {\it et al.} (CDF Collaboration), Phys. Rev. Lett.
{\bf 86}, 4472 (2001); 
V.M. Abazov {\it et al.} (D0 Collaboration), Phys. Rev. Lett. {\bf 97}, 171806 (2006).

\bibitem{bsgamma} E. Barberio {\em et al.} (HFAG), arXiv:hep-ex/0603003.


\bibitem{bsgamma_theorySUSY}
M. Ciuchini, G. Degrassi, P. Gambino and G.F. Giudice, Nucl. Phys. B {\bf 534}, 3 (1998).

\bibitem{bsgamma_theorySM}
M. Misiak {\em et al.}, Phys. Rev. Lett. {\bf 98}, 022002 (2007).

\bibitem{bsmumu} V.M. Abazov {\em et al.}, (D0 Collaboration), 
Phys. Rev.  D {\bf 76}, 092001 (2007).

\bibitem{bennet} G.W. Bennet at al. (Muon g-2 Collaboration), Phys. Rev. D {\bf} 73, 072003 (2006).

\bibitem{bijnens} J. Bijnens and J. Prades,  
Mod. Phys. Lett.  A {\bf 22}, 767 (2007).


\bibitem{esa} The first of Eqs. (\ref{diag}) is also derived
in M. M. El Kheishen, A. A. Shafik and A. A. Aboshousha, Phys. Rev. D {\bf 67}, 4345
(1992).

\bibitem{wmapetc} D.N. Spergel {\it et al.}, Astrophys. J. Suppl. Ser.
{\bf 148}, 175 (2003); M. Tegmark {\it et al.}, Phys. Rev. D {\bf 69}, 103501
(2004).

\bibitem{others} This lower limit is also found in D. Hooper and T. Plehn, Phys. Lett. {\bf B562} (2003) 18
and  G. Belanger, F. Boudjema, A. Pukhov andS. Rosier-Lees, arXiv:hep-ph/0212227.




\bibitem{Bachacou:2000zb}
H.~Bachacou, I.~Hinchliffe and F.~E.~Paige,
Phys.\ Rev.\ D {\bf 62}, 015009 (2000).

\bibitem{Allanach:2000kt}
B.~C.~Allanach, C.~G.~Lester, M.~A.~Parker and B.~R.~Webber,
JHEP{\bf 0009}, 004 (2000).

\bibitem{Gjelsten:2004ki}
  B.~K.~Gjelsten, D.~J.~Miller and P.~Osland,
  JHEP {\bf 0412} (2004) 003.

\bibitem{Nojiri:2005ph}
M.~M.~Nojiri, G.~Polesello and D.~R.~Tovey,
JHEP {\bf 0603}, 063 (2006).


\bibitem{cinque} A.~Datta, A.~Djouadi, M.~Guchait and Y.~Mambrini,
Phys.\ Rev.\  D {\bf 65}, 015007 (2002); K.~Huitu, J.~Laamanen, P.~N.~Pandita and S.~Roy,
Phys.\ Rev.\  D {\bf 72}, 055013 (2005).

\bibitem{isasusy} 
  F.~E.~Paige, S.~D.~Protopopescu, H.~Baer and X.~Tata,
  arXiv:hep-ph/0312045.

\bibitem{prospino}
W.~Beenakker, R.~Hopker and M.~Spira,
  arXiv:hep-ph/9611232;\\
W.~Beenakker, R.~Hopker, M.~Spira and P.~M.~Zerwas,
  Nucl.\ Phys.\  B {\bf 492} (1997) 51.
  
 \bibitem{future} A. Bottino, N. Fornengo, G. Polesello and S. Scopel  
 (to appear). 
  
\end{thebibliography}
\end{document}